\documentclass{article}

\usepackage{graphicx}
\usepackage{amsmath}
\usepackage{amssymb}
\usepackage{calc}
\usepackage{geometry}
\usepackage{units}

\newcommand{\confidence}{\ensuremath{{\mathrm{C}}}}
\newcommand{\emag}{\ensuremath{\mathrm{em}}}

\newcommand{\LEQ}{\ensuremath{\mathcal{L}_{\mathrm{EQ}}}}

\newcommand{\LNC}{\ensuremath{\mathcal{L}_{\mathrm{NC}}}}
\newcommand{\alphaem}{\ensuremath{\alpha_\mathrm{em}}}

\newcommand{\threesigma}{\ensuremath{3\sigma}}
\newcommand{\fivesigma}{\ensuremath{5\sigma}}

\newcommand{\group}[1]{\ensuremath{\mathrm{#1}}}
\newcommand{\Dbar}{\overline}
\newcommand{\SUtwo}{\group{SU(2)}}
\newcommand{\Uone}{\group{U(1)}}
\newcommand{\SUN}{\group{SU(\mathnormal{N})}}

\newcommand{\FNAL}{FNAL}
\newcommand{\CDF}{CDF}
\newcommand{\Dzero}{D\O{}}
\newcommand{\LEP}{LEP}
\newcommand{\LHC}{LHC}
\newcommand{\ALEPH}{ALEPH}
\newcommand{\OPAL}{OPAL}
\newcommand{\tautau}{\tauplus\tauminus}
\newcommand{\pythia}{\textsc{Pythia}}
\newcommand{\pt}{\ensuremath{{p_T}}}
\newcommand{\ptl}{\ensuremath{{p_T^L}}}
\newcommand{\pte}{\ensuremath{{p_T^\elec}}}
\newcommand{\ptmu}{\ensuremath{{p_T^\muon}}}
\newcommand{\pthresh}{\ensuremath{{P_T^{\text{cut}}}}}
\newcommand{\met}{\ensuremath{\not E}}
\newcommand{\mett}{\ensuremath{\met_T}}
\newcommand{\TEVtwoK}{TeV\ 2000}

\newcommand{\particle}[1]{\ensuremath{\mathrm{#1}}}

\newcommand{\proton}{\ensuremath{\mathrm{p}}}

\newcommand{\quark}{\ensuremath{\mathrm{q}}}
\newcommand{\quarkup}{\ensuremath{\mathrm{u}}}
\newcommand{\qup}{\quarkup}
\newcommand{\quarkdown}{\ensuremath{\mathrm{d}}}
\newcommand{\qdown}{\quarkdown}
\newcommand{\quarkcharm}{\ensuremath{\mathrm{c}}}
\newcommand{\qcharm}{\quarkcharm}
\newcommand{\quarkstrange}{\ensuremath{\mathrm{s}}}
\newcommand{\qstrange}{\quarkstrange}
\newcommand{\quarktop}{\ensuremath{\mathrm{t}}}
\newcommand{\qtop}{\quarktop}
\newcommand{\quarkbottom}{\ensuremath{\mathrm{b}}}
\newcommand{\qbottom}{\quarkbottom}

\newcommand{\neutrino}{\ensuremath{\mathrm{\nu}}}

\newcommand{\electron}{\ensuremath{\mathrm{e}}}
\newcommand{\elec}{\electron}
\newcommand{\eplus}{\ensuremath{\electron^+}}
\newcommand{\eminus}{\ensuremath{\electron^-}}
\newcommand{\nue}{\ensuremath{\mathrm{\nu}_\electron}}

\newcommand{\muon}{\ensuremath{\mathrm{\mu}}}
\newcommand{\muplus}{\ensuremath{\muon^+}}
\newcommand{\muminus}{\ensuremath{\muon^-}}

\newcommand{\tauon}{\ensuremath{\mathrm{\tau}}}
\newcommand{\tauplus}{\ensuremath{\mathrm{\tau}^+}}
\newcommand{\tauminus}{\ensuremath{\mathrm{\tau}^-}}

\newcommand{\photonA}{\ensuremath{\mathrm{A}}}

\newcommand{\Wparticle}{\ensuremath{\mathrm{W}}}
\newcommand{\Wpart}{\Wparticle}

\newcommand{\Wplus}{\ensuremath{{\mathrm{W}^+}}}
\newcommand{\Wminus}{\ensuremath{{\mathrm{W}^-}}}

\newcommand{\Wprime}{\ensuremath{{\mathrm{W}'}}}

\newcommand{\Znaught}{\ensuremath{\mathrm{Z}^0}}
\newcommand{\Zprime}{\ensuremath{{\mathrm{Z}'}}}

\makeatletter
\@addtoreset{equation}{section}
\makeatother

\newcommand{\bu}{$^1$}
\newcommand{\ucd}{$^2$}
\newcommand{\addlabel}[2]{\thanks{#1}\ $^{\text{,}}${#2}}

\title{Finding \Zprime{} Bosons Coupled Preferentially to the Third
  Family \\ at CERN LEP and the Fermilab Tevatron}
\author{Kevin R. Lynch\addlabel{krlynch@bu.edu}{\bu},
  Stephen Mrenna\addlabel{mrenna@physics.ucdavis.edu}{\ucd},
  Meenakshi Narain\addlabel{narain@bu.edu}{\bu}, \and
  Elizabeth H. Simmons\addlabel{simmons@bu.edu}{\bu}\\
  \\
  \bu\ Department of Physics, Boston University, \\
  590 Commonwealth Avenue, Boston MA  02215\\
  \\
  \ucd\ Physics Department, University of California, Davis\\
  One Shields Avenue, Davis CA  95616}
\date{July 25, 2000}

\begin{document}

\begin{titlepage} 
\maketitle
\thispagestyle{empty}
 
\bigskip
\begin{picture}(0,0)(0,0)
\put(400,300){BUHEP-00-4}
\put(400,285){UCD-2000-12}
\put(400,270){hep-ph/0007286}
\end{picture}
\vspace{24pt}
 
\begin{abstract}
  
  \Zprime{} bosons that couple preferentially to the third generation
  fermions can arise in models with extended weak ($\SUtwo\times
  \SUtwo$) or hypercharge ($\Uone\times \Uone$) gauge groups.  We show
  that existing limits on quark-lepton compositeness set by the LEP
  and Tevatron experiments translate into lower bounds of order a few
  hundred $\unit{GeV}$ on the masses of these \Zprime{} bosons.
  Resonances of this mass can be directly produced at the Tevatron.
  Accordingly, we explore in detail the limits that can be set at Run
  II using the process $\proton \Dbar{\proton} \rightarrow \Zprime{}
  \rightarrow \tauon \tauon \rightarrow \elec \muon$.  We also comment
  on the possibility of using hadronically-decaying taus to improve
  the limits.
\end{abstract}
\end{titlepage}


\section{Introduction}
\label{sec:intro}

The Standard Model of particle physics gives an excellent description
of physics at the energy scales probed to date.  Nonetheless, it does
not explain the origins of the masses of the electroweak gauge bosons
and the elementary fermions, and must be regarded as a low-energy
effective field theory.  For a description of the dynamics underlying
the generation of mass, we must turn to physics beyond the Standard
Model.

Much recent theoretical work on the question of why the top quark is
so heavy has suggested that the cause could be additional gauge
interactions that single out the third generation fermions. A number
of interesting models along these lines extend one (or more) of the
Standard Model's \SUN{} gauge groups into an $\SUN \times \SUN$
gauge structure \cite{topcolor, ncETC, topassist, coloron,
topflavor}. In general, fermions of the third generation transform
under one \SUN{} group and those of the first and second generations
transform under the other one.  When the $\SUN\times \SUN$
spontaneously breaks to its diagonal subgroup, the broken generators
correspond to a set of massive \SUN{} gauge bosons that couple to
fermions of different generations with different strengths.

Many of these models predict the presence of massive \Zprime{} bosons
that couple preferentially to the third-generation fermions.  Some
theories include an extended $\SUtwo \times \SUtwo$ structure for the
weak interactions; generally, the first two generations of fermions
are charged under the weaker \SUtwo{} and the third generation feels
the other, stronger, \SUtwo{} gauge force.  Examples include
non-commuting extended technicolor (NCETC) models \cite{ncETC,
singletop} and topflavor models \cite{topflavor, topflavses}.  For
many of these models, precision data suggest that the \Zprime{} must
be relatively heavy; but in some non-commuting extended technicolor
models, a mass as low as $\unit[400]{GeV}$ is not precluded
\cite{ncETC}.  There are also theories in which a \Zprime{} boson
arises due to an extra \Uone{} group coupled preferentially to the
third-generation quarks (and possibly leptons).  Examples are the
topcolor-assisted technicolor \cite{topassist, topastmod} and
flavor-universal topcolor-assisted-technicolor \cite{KDL, zpcol}
models.  In these models, depending on the charge assignments of the
ordinary and technifermions, constraints from FCNC and precision
electroweak corrections can allow the \Zprime{} boson to be as light
as $\unit[290]{GeV}$ \cite{rscjt}.

More generally, electroweak scale \Zprime\ bosons are also present in
string theories \cite{pgla}, and string-inspired models often yield
non-universal couplings \cite{pglb} for the \Zprime.  A recent
analysis of electroweak precision data \cite{pglc} actually gives a
strong indication of the presence of an extra \Zprime\ boson with the
fits favoring non-universal couplings to the third family.

The literature already contains a number of suggestions about how
experiment can set stronger limits on these \Zprime{} bosons.  Note
that bounds on \Zprime{} bosons which do not couple preferentially to
the third generation \cite{godfrey} are not directly applicable.  For
example, in models with extended weak interactions, the presence of
\Wprime{} bosons with mass below about \unit[1.5]{TeV} would cause an
enhancement of single top quark production large enough to be visible
at the Tevatron's Run II experiments \cite{singletop}; this would
provide indirect evidence of similarly light \Zprime{} bosons.  The
Run I Tevatron experiments have searched for topcolor \Zprime\ bosons
in $\qbottom\Dbar{\qbottom}$ and $\qtop\Dbar{\qtop}$ final states.  In
these processes, the backgrounds are of QCD strength.  As a result, no
limit has been set from the $\qbottom\Dbar{\qbottom}$ channel
\cite{cdftopcolor} and the recent limit in the $t\bar{t}$ channel
($M_{\Zprime} > \unit[650]{GeV}$) \cite{cdfttbar} is for a \Zprime\
that couples only to hadrons and is quite narrow, $\Gamma_{\Zprime} =
.012 M_{\Zprime}$.  These searches should have greater reach in Run
II, due to the higher luminosity and improved detectors.
Flavor-changing neutral current effects can also yield constraints on
\Zprime\ bosons with non-universal couplings \cite{pgld}.

This paper discusses two additional methods of searching for \Zprime{}
bosons that couple primarily to the third family fermions.  We first
show how existing \LEP{} and Tevatron bounds on the scale of
quark-lepton compositeness can be adapted to provide lower bounds on
the mass of these \Zprime{} bosons.  We then analyze, the possibility of
searching at Tevatron Run II for \Zprime{} bosons in the channel
$\proton\Dbar{\proton} \rightarrow \Zprime \rightarrow \tauon\tauon
\rightarrow \elec\muon$ in order to exploit the strong
$\Zprime\to\tauon\tauon$ coupling and the low backgrounds for
$\elec\muon$ final states. 

In Section~\ref{sec:ewsb}, we will review the properties of the
\Zprime{} boson arising in models with an extended weak gauge group
and display the existing limits from electroweak precision data.  In
Section~\ref{sec:composite} we extract limits on these \Zprime{}
bosons from the \LEP{} and Tevatron compositeness bounds.
Section~\ref{sec:channels} focuses on the Run II search in
leptonically-decaying pair-produced taus.  We then in
Section~\ref{sec:exthyp} show how our results are modified for
\Zprime{} bosons in models with an extended hypercharge group, and we
mention a few additional search channels which may help improve the
reach of Run II in Section~\ref{sec:disc}.  Our conclusions are
presented in Section~\ref{sec:conclusions}.


\section{\Zprime{} Bosons From Extended Weak Interactions}
\label{sec:ewsb}

\subsection{General properties of \SUtwo{} \Zprime{} bosons}

The models of interest to us include the usual complement of quarks and
leptons, along with standard strong and hypercharge interactions.  The
new physics lies in the weak interactions, which are governed by a pair
of \SUtwo{} gauge groups:
\begin{equation}
 \SUtwo_h \times \SUtwo_\ell\ .
\label{eqn:thegrp}
\end{equation}
The $\SUtwo_h$ group governs the weak interactions for the third
generation (heavy) fermions; the left-handed fermions transform as doublets
and the right-handed ones, as singlets under this group.  Similarly,
the $\SUtwo_\ell$ group couples to the first and second
generation (light) fermions, whose charges under $\SUtwo_\ell$ are as in the
standard model.  The extended weak group (Equation~\ref{eqn:thegrp}) is broken
to its diagonal subgroup, $\SUtwo_L$ at energy scale $u$ by a
(composite) scalar field $\sigma$, charged under 
$\SUtwo_h \times \SUtwo_\ell \times \Uone_Y$ as:
\begin{alignat}{2}
\sigma &\sim (2,2)_0, & \qquad
\left< \sigma \right> & = \left(\begin{array}{cc}u & 0 \\ 0 & u
\end{array} \right)\ . 
\end{alignat} 

The final step in electroweak symmetry breaking could, in principle,
proceed through a condensate charged under $\SUtwo_\ell$ or one
charged under $\SUtwo_h$ -- or one of each \cite{ncETC, topflavor}.
The first option allows the \Zprime{} boson to be 
the lightest \cite{ncETC}, making it the option of
greatest phenomenological interest.  Hence, we assume that the
symmetry breaking $\SUtwo_L \times \Uone_Y \rightarrow \Uone_{\emag}$ is due 
to a
(composite) scalar
\begin{alignat}{2}
\Phi & \sim (1,2)_{1/2}, &\qquad
\left< \Phi \right> & = \left(\begin{array}{c} 0 \\ v/\sqrt{2}\end{array}
\right)\ .
\end{alignat}

The generator of the
$\Uone_{\emag}$ group is the electric charge operator:
\begin{equation}
Q = T_{3h} + T_{3\ell} + Y\ ,
\label{eqn:Q}
\end{equation}
and the corresponding photon eigenstate is:
\begin{equation}
\photonA^\mu = \sin\theta \left( \cos\phi \,\Wpart^\mu_{3h} + \sin\phi 
\,\Wpart^\mu_{3\ell}
\right) + \cos\theta \,\particle{X}^\mu
\label{eqn:photon}
\end{equation}
where $\theta$ is the usual weak mixing angle and $\phi$ is an additional 
mixing angle occasioned by the presence of two weak gauge groups.
We can therefore relate the gauge couplings and mixing angles as follows:
\begin{equation}
\begin{gathered}
g_h = \frac{e}{\cos\phi\, \sin\theta} = \frac{g}{\cos\phi} \\
g_\ell = \frac{e}{\sin\phi\, \sin\theta} = \frac{g}{\sin\phi} \\
g_Y = \frac{e}{\cos\theta}\ .
\label{eqn:coupgath}
\end{gathered}
\end{equation}
For brevity we will write $s_\phi \equiv \sin\phi$ and
$c_\phi \equiv \cos\phi$.

In diagonalizing the mass matrix for the neutral gauge bosons, it is
convenient to first transform to an intermediate basis
\cite{ununified},
\begin{gather}
\particle{Z}_1^\mu = \cos\theta \left(c_\phi \Wpart_{3h}^\mu + s_\phi 
\Wpart_{3\ell}^\mu\right) -
\sin\theta \,\particle{X}^\mu \label{eqn:Z1} \\
\particle{Z}_2^\mu = -s_\phi \Wpart_{3h}^\mu + c_\phi \Wpart_{3\ell}^\mu 
\label{eqn:Z2} \ ,
\end{gather}
where the covariant derivative neatly separates into
standard and non-standard contributions:
\begin{equation} 
D^\mu = \partial^\mu -i\frac{g}{\cos\theta} \particle{Z}^\mu_1
\left(T_{3h}+T_{3\ell}-\sin^2\theta \,Q\right) - ig \particle{Z}^\mu_2
  \left(-\frac{s_\phi}{c_\phi} T_{3h} + \frac{c_\phi}{s_\phi} T_{3\ell}
  \right)\ .
\label{eqn:covariant}
\end{equation}
In terms of these states, the neutral mass eigenstates (the \Znaught{} and
\Zprime{} states) are given, to leading order in $1/x = v^2/u^2$,  by the
superpositions:
\begin{equation}
\left(\begin{array}{c}\Znaught \\ \Zprime\end{array}\right) \approx
  \left(\begin{array}{cc} 1 & -\frac{c_\phi^3 s_\phi}{x \cos\theta} \\
    \frac{c_\phi^3 s_\phi}{x \cos\theta} & 1\end{array}\right) \left(
    \begin{array}{c} \particle{Z}_1 \\ \particle{Z}_2\end{array}\right)\ .
\end{equation}

Expanding the covariant derivative in Equation~\ref{eqn:covariant} in terms
of the mass eigenstates \Znaught\ and \Zprime, we find that, to order $1/x$:
\begin{multline}
  D_\mu = \partial_\mu - \frac{ig}{\cos\theta} \Znaught_\mu \left[ \left( 1 -
  \frac{c^4_\phi}{x} \right) T_{3\ell} + \left(1+\frac{c^2_\phi
    s^2_\phi}{x} \right) T_{3h} -\sin^2\theta \,Q \right]
  \label{eqn:covariant_b}\\  
-ig\Zprime_\mu \left[\left(\frac{c_\phi}{s_\phi} + \frac{c^3_\phi s_\phi}{x
  \cos^2\theta}\right) T_{3\ell} + \left(-\frac{s_\phi}{c_\phi} +
\frac{c^3_\phi s_\phi}{x \cos^2\theta}\right) T_{3h} - \sin^2\theta
\left(\frac{c^3_\phi s_\phi}{x \cos^2\theta}\right) Q\right] \ .
\end{multline}
For large $s_\phi$ the \Znaught{} boson can maintain a nearly standard
model coupling to all fermion species, while the \Zprime{} boson has a
greatly enhanced coupling to the third generation fermions.  Moreover,
we will see that the $1/x$ corrections are small in the
phenomenologically interesting region of parameter space, so that the
\Zprime{} boson essentially couples only to left-handed fermions.

To leading order, the mass of the \Zprime{} boson in the region where $s_\phi$ 
exceeds $c_\phi$ is
\begin{equation}
  M^2_{\Zprime} = \left(\frac{ev}{2\sin\theta}\right)^2
    \frac{x}{s^2_\phi c^2_\phi} = M^2_{\Wpart_{\text{SM, tree}}}
    \frac{x}{s^2_\phi c^2_\phi}\ .
\label{eqn:zzmass}
\end{equation}
and, to this order, the mass of the \Wprime{} boson is the
same.  The masses of the
\Znaught{} and $W^\pm$ bosons are shifted from their tree-level
standard model values by identical multiplicative factors, so that
there is no change in the predicted value of the $\rho$ parameter at
order $1/x$ \cite{ncETC}.

\begin{figure}[tb]
\begin{center}
\includegraphics[width=\textwidth-2in]{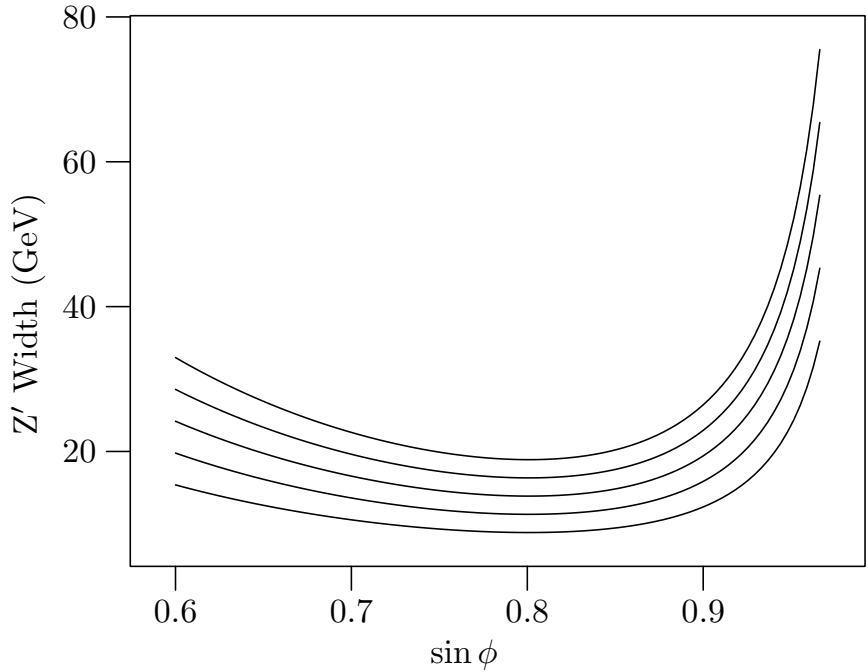}
\end{center}
\caption{\Zprime{} widths for five different \Zprime{} masses (from
  below: \unit[350]{GeV}, \unit[450]{GeV}, \unit[550]{GeV},
  \unit[650]{GeV}, and \unit[750]{GeV}) as a function of the
  mixing angle $s_\phi$.  Note that as $s_\phi$ increases, the width
  of the \Zprime{} falls to a minimum in the neighborhood of $s_\phi =
  0.8$, due to the decreasing couplings to the first and second
  generation fermions.  The width then rises rapidly as $s_\phi$ grows
  large, due to the rapid growth in the third generation couplings.}
\label{fig:widths}
\end{figure}

The width of the \Zprime, to leading (here, zeroth) order in $1/x$  and in
the region where $s_\phi > c_\phi$ is\footnote{A fermionic species, $f$, 
contributes to the width of this gauge boson as
\begin{equation}
\Gamma_{f,\Zprime} = C_f \frac{\left(M_{\Zprime}^2 - 4
    m_f^2\right)^{1/2}}{48\pi} \left[\left(g_{f,R} +
    g_{f,L}\right)^2\left(1 + \frac{2 m_f^2}{M_{\Zprime}^2}\right) +
  \left(g_{f,R} - g_{f,L}\right)^2\left(1 - \frac{4
      m_f^2}{M_{\Zprime}^2}\right)\right]\, ,  
\end{equation}
where $C_f$ is a color factor (1 for leptons, 3 for quarks), $m_f$
is the fermion mass, and $g_{f,R}$, $g_{f,L}$ are the right and left
handed couplings of the fermion.}
\begin{equation}
  \frac{\Gamma_\Zprime}{\alpha_2 M_\Zprime} =
  \frac{2}{3}\left(\frac{c_\phi}{s_\phi}\right)^2 +
  \left[\frac{5}{24} + \frac{1}{8}
  \left(1 - \frac{2 m_\qtop^2}{M_{\Zprime}^2}\right)\left(1 - \frac{4
      m_\qtop^2}{M_{\Zprime}^2}\right)^{1/2} \Theta(M_{\Zprime}-2
  m_\qtop)\right]   \left(\frac{s_\phi}{c_\phi}\right)^2 \ ,
\end{equation}
where we have included the effects of the \qtop-quark mass, while
taking the other fermion masses to be zero.  The step function ensures
that we only include the top contribution when the \Zprime{} mass is
above the top threshold.  To this order, the \Zprime{} boson couples
neither to right-handed fermions nor to \Znaught\Znaught{} nor to
\Wplus\Wminus{}.  Effects from the composite scalars are not
substantial.  Figure~\ref{fig:widths} shows the \Zprime{} width for
\Zprime{} masses between \unit[350]{GeV} and
\unit[750]{GeV} as a function of the mixing angle $s_\phi$.
The width of the \Zprime{} falls to a minimum at approximately $s_\phi
= 0.8$.  The width then grows rapidly as $s_\phi$ becomes larger.

\subsection{Light \SUtwo{} \Zprime{} in models of dynamical symmetry breaking}

\begin{figure}[tb]
\begin{center}
\includegraphics[width=\textwidth-2in]{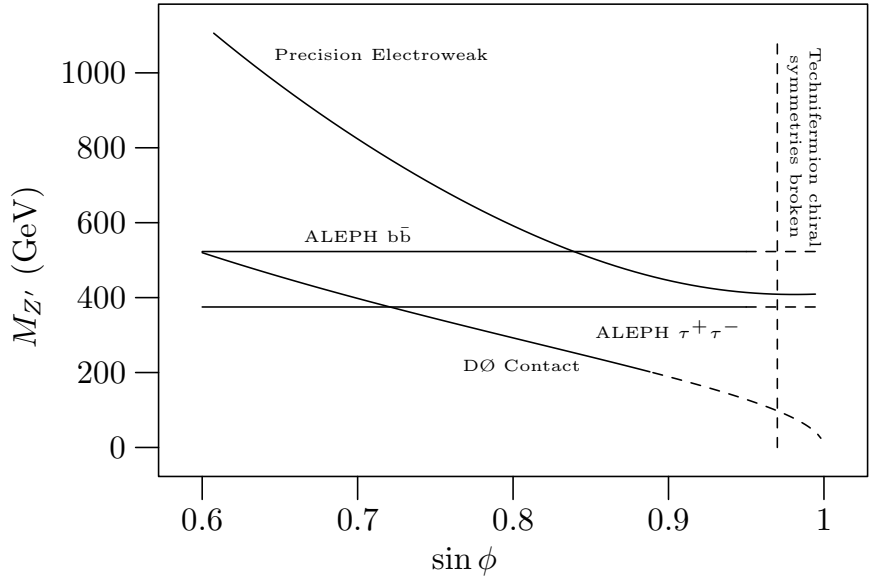}
\end{center}
\caption{ Limits at the 95\% confidence level on the
  mass of the \Wprime{} and \Zprime{} bosons (which are equal to
  leading order in $1/x$), as a function of the parameter $s_\phi$ from
  various sources.  Regions of the parameter space lying below a given
  line are excluded by the corresponding search.  The curve from upper
  left to lower right is based on precision electroweak data \cite{ncETC}.  
  The horizontal lines are the current limits from the compositeness analyses 
  at \LEP{}; as $s_\phi$ approaches 1, our
  approximations may break down due to the rapid increase in the
  \Zprime{} width.  The bottom curve is the current  limit
  from the compositeness analyses at \FNAL{} (first generation
  fermions at \Dzero{}, Section~\ref{sec:fnaldata}); the contact approximation 
  breaks down for light \Zprime.  The region to the right of the vertical 
  dashed line is excluded to avoid $\SUtwo_h$ couplings strong enough to 
  break the chiral symmetries of the
  technifermions.}
\label{fig:ew-comp-limits}
\end{figure}

A gauge structure of this kind has been proposed within the context of
models of dynamical electroweak symmetry breaking \cite{ncETC,
topflavses} and models in which the vacuum expectation value of a
weakly coupled scalar boson breaks the electroweak symmetry
\cite{topflavor}.  A class of models in which the precision
electroweak data allow the \Zprime{} and \Wprime{} bosons to be
particularly light are the non-commuting extended technicolor models
\cite{ncETC}.  The electroweak symmetry breaking pattern of
non-commuting extended technicolor is characterized by a three-stage
breakdown from the unbroken, high energy theory to the low energy
electromagnetic gauge structure:
\begin{equation}
\begin{split}
\group{G}_{ETC} \times \SUtwo_\ell \times \Uone' &
\stackrel{f}{\longrightarrow}\\ 
\group{G}_{TC} \times \SUtwo_h \times \SUtwo_\ell
\times \Uone_Y & \stackrel{u}{\longrightarrow} \\
\group{G}_{TC} \times \SUtwo_L \times \Uone_Y &
\stackrel{v}{\longrightarrow}\\ 
\group{G}_{TC} \times \Uone_{\emag} & \ .
\end{split}
\end{equation}
At the scale $f$, the extended technicolor gauge group
$\group{G}_{ETC}$ breaks to the technicolor gauge group and the
$\SUtwo_h$ (heavy) group.  The two $\SUtwo$ groups mix and break to
their diagonal subgroup at the scale $u$.  The final breaking of the
remaining electroweak symmetry is accomplished at scale $v$.  If the
condensate $\langle\Phi\rangle$ responsible for electroweak symmetry
breaking at scale $v$ is charged under $\SUtwo_\ell$ (rather than
$\SUtwo_h$), the resulting masses of the \Zprime{} and \Wprime{}
bosons can be as low as \unit[400]{GeV}, as illustrated in
Figure~\ref{fig:ew-comp-limits}. 
New gauge bosons with such small masses are of great phenomenological
interest, as they are within the kinematic reach of Tevatron Run II
experiments and their indirect effects may be apparent at \LEP{} 2.
In other models with the extended $\SUtwo_h \times \SUtwo_\ell$ gauge
structure, existing lower limits on the gauge boson masses tend to be
of order \unit[1 - 1.5]{TeV} \cite{topflavor}.

Note that in the context of non-commuting extended technicolor models,
the coupling $g_h$ is essentially the value of the technicolor
coupling at scale $f$.  We therefore expect $g_h$ to be large compared
to the weak coupling $g$, so that the value of $c^2_\phi$ (from
Equation~\ref{eqn:coupgath}) should be relatively small.  However, if
$c^2_\phi$ is too small, $g_h$ will be above the critical value at
which the chiral symmetries of the technifermions break.  Thus, as
discussed in \cite{ncETC}, we must restrict $s^2_\phi$ to be smaller
than about 0.95, hence $s_\phi \lesssim 0.975$ (vertical dashed line in
Figure~\ref{fig:ew-comp-limits}).


\section{Limits on an \SUtwo{} \Zprime{} from Compositeness Searches}
\label{sec:composite}

For experiments done at energies below the mass of a \Zprime{} boson,
one can approximate the contribution of the \Zprime{} to fermion-fermion
scattering as a contact interaction whose scale is set by the
mass of the \Zprime{} boson.  Thus, published experimental limits on
compositeness can set a lower bound on $M_{\Zprime}$.

\subsection{\LEP{} Data}
\label{sec:lepdata}

The \LEP{} experiments \ALEPH{} \cite{aleph-183} and \OPAL{} \cite{opal-183} 
and have recently published limits on contact interactions.
Following the notation of \cite{elp}, they write the effective Lagrangian for
the four-fermion contact interaction in the process $\eplus \eminus 
\rightarrow f
\Dbar{f}$ as
\begin{equation} 
\mathcal{L}_{\text{contact}} = \frac{g^2}{\Lambda^2 (1 + \delta)}
\sum_{i,j = L, R} \eta_{ij} \big( \Dbar{\elec}_i \gamma_\mu
\elec_i\big) \big(\Dbar{f}_j \gamma^\mu f_j\big)
\end{equation}
where $\delta = 1$ if $f$ is an electron and $\delta = 0$ otherwise.
The values of the coefficients $\eta_{ij}$ set the chirality structure
of the interaction being studied; \OPAL{} and \ALEPH{} study a number of
cases where one of the $\eta_{ij}$ is equal to $\pm 1$ and the others
are zero.  Following the convention \cite{elp} of taking $g^2 / 4\pi = 1$,
they determine a lower bound on the scale $\Lambda$ associated with
each type of new physics.  Of particular interest to us are their
limits on contact interactions where the final-state fermions $f$
belong to the third generation: $\eplus \eminus \rightarrow \qbottom
\Dbar{\qbottom}$ and $\eplus \eminus
\rightarrow \tauplus \tauminus$.  Among the limits published by \ALEPH{}
\cite{aleph-183} and \OPAL{} \cite{opal-183}, those of interest
to us are
\begin{align}
\Lambda(f = \tauon,\ \eta_{LL} = +1) & >\begin{cases} \unit[3.9]{TeV} &
  \text{\ALEPH}\\ \unit[3.8]{TeV} & \text{\OPAL}\end{cases}\\
\Lambda(f = \qbottom,\ \eta_{LL} = +1) & >\begin{cases}  \unit[5.6]{TeV} &
  \text{\ALEPH}\\ \unit[4.0]{TeV} & \text{\OPAL}\ .\end{cases}
\end{align}

At energies well below the mass of the \Zprime{} boson, its exchange
in the process $\eplus\eminus \rightarrow f \Dbar{f}$ where $f$ is
a \tauon{} lepton or 
\qbottom{} quark may be approximated by the contact interaction
\begin{equation} 
\LNC \supset -\frac{e^2}{\sin^2\theta\, M^2_{\Zprime}}
\left(-\frac{c_\phi}{2 s_\phi} \big(\Dbar{\elec}_L \gamma_\mu
  \elec_L\big)\right) \left(\frac{s_\phi}{2 c_\phi}
  \big(\Dbar{f}_L \gamma^\mu f_L\big)\right)\ , 
\end{equation}
based on the \Zprime{}-fermion couplings in
Equation~\ref{eqn:covariant_b}.  Comparing this with the contact
interactions studied by \LEP{}, we find
\begin{equation}
M_{\Zprime} = \Lambda \sqrt{\frac{\alphaem}{4 \sin^2\theta}}\ .
\label{zplime}
\end{equation}
The limits from \tauon{}-pair production are, then,
\begin{equation}
M_{\Zprime} >\begin{cases} \unit[365]{GeV} & \text{\ALEPH}\\
  \unit[355]{GeV} & \text{\OPAL}\ ,\end{cases}
\end{equation}
and those from $\qbottom\Dbar{\qbottom}$ production are
\begin{equation}
M_{\Zprime} >\begin{cases} \unit[523]{GeV} & \text{\ALEPH}\\
  \unit[375]{GeV} & \text{\OPAL}\ .\end{cases}
\end{equation}
As Figure~\ref{fig:ew-comp-limits} illustrates, the limits are
comparable to the previous lower bound on $M_{\Zprime}$ from precision
electroweak data in the case where $s_\phi$ is large; for small
$s_\phi$ the earlier limits remain stronger.  As additional data from
the other experiments or higher energies becomes available, the lower
bound on $M_{\Zprime}$ can be updated by using the new lower bound on
$\Lambda$ in Equation~\ref{zplime}.

\subsection{\FNAL{} Data}
\label{sec:fnaldata}

The \CDF{} and \Dzero{} Collaborations have each searched for the low
energy effects of quark-lepton contact interactions on dilepton
production in $\unit[110]{pb^{-1}}$ of data taken at $\sqrt{s} =
\unit[1.8]{TeV}$ \cite{cdf-dylim}.  Since this process is dominated by
first and second generation fermions, the limits on our \Zprime{}
bosons tend to be weaker than those derived from \LEP{} data.

In their analysis, the \CDF{} Collaboration described the
effective four-fermi interactions of the first generation fermions due
to new physics by an effective Lagrangian including the terms:
\begin{multline}
\LEQ \supset \xi^0_{LL} \big(\Dbar{\particle{E}}_L \gamma_\mu \particle{E}_L
\big)\big( \Dbar{\particle{Q}}_L \gamma^\mu \particle{Q}_L\big)
\,+\, \xi^1_{LL} \big(\Dbar{\particle{E}}_L \gamma_\mu \tau_a
\particle{E}_L\big)\big(\Dbar{\particle{Q}}_L \gamma^\mu \tau_a 
\particle{Q}_L\big) \\
\,+\, \xi^u_{LR} \big(\Dbar{\particle{E}}_L \gamma_\mu \particle{E}_L\big)
\big(\Dbar{\qup}_R \gamma^\mu \qup_R\big)
\,+\, \xi^d_{LR} \big(\Dbar{\particle{E}}_L \gamma_\mu \particle{E}_L\big)
\big(\Dbar{\qdown}_R \gamma^\mu \qdown_R\big)
\,+\, \xi^e_{RL} \big(\Dbar{\elec}_R \gamma_\mu \elec_R\big)
\big(\Dbar{\particle{Q}}_L \gamma^\mu \particle{Q}_L\big)\\
\,+\, \xi^u_{RR} \big(\Dbar{\elec}_R \gamma_\mu \elec_R\big)
\big(\Dbar{\qup}_R \gamma^\mu \qup_R\big) 
\,+\, \xi^d_{RR} \big(\Dbar{\elec}_R \gamma_\mu \elec_R\big) 
\big(\Dbar{\qdown}_R \gamma^\mu \qdown_R\big)\ ,
\label{eqn:LEQ}
\end{multline}
where $\particle{Q}_L \equiv \left(\qup,\qdown\right)_L$,
$\particle{E}_L \equiv \left(\nue,\elec\right)$, and the subscripts
$L$ and $R$ denote the left and right helicity projections.  The
coefficients $\xi_{ij}$ are related to the scale of new physics,
$\Lambda_{ij}$, as $\xi_{ij} = g^2_0 / \Lambda_{ij}$, where $g^2_0$ is
an effective coupling which grows strong at the compositeness scale:
${g^2_0(\Lambda)}/{4\pi} = 1$.  The analysis searched for
deviations in the dilepton spectrum from the standard model
prediction; the absence of such deviations enabled them to set a lower
bound on the scale of the new interactions.

The \CDF{} analysis included fermions beyond the first generation by
assuming a kind of universality: electrons and muons have identical
contact interactions, all up-type quarks behave alike, all down-type
quarks behave alike.  They derived separate limits on contact
interactions involving different combinations of fermions; for example,
assuming that the only contact interaction was one between left-handed
muons (electrons) and up-type quarks, they found at $95\%$ confidence
\begin{align}
\Lambda(\muon_L; \qup_L, \qcharm_L, \qtop_L) & > \unit[4.1]{TeV}\\
\Lambda(\elec_L; \qup_L,\qcharm_L, \qtop_L) & > \unit[3.7]{TeV}\ .
\label{eqn:cdflimit}
\end{align}

The presence of a massive \Zprime{} boson in our model gives rise to
four-fermion contact interactions that include the terms
\begin{equation} 
\LNC \supset -\frac{e^2}{\sin^2\theta M^2_{\Zprime}}
\left(\frac{c_\phi}{2 s_\phi}\right)^2 
\big(\Dbar{\elec}_L \gamma_\mu \elec_L + \Dbar{\muon}_L \gamma_\mu \muon_L\big)
\big(\Dbar{\qup}_L \gamma^\mu \qup_L + \Dbar{\qdown}_L \gamma^\mu \qdown_L 
+\Dbar{\qcharm}_L \gamma^\mu \qcharm_L + \Dbar{\qstrange}_L \gamma^\mu 
\qstrange_L \big)\ .
\label{eqn:ourzp}
\end{equation}
In other words, the contact interactions among left-handed fermions in
the first two generations all have the same coefficient.  The
interaction strength for third-generation fermions is different, as
seen from Equation~\ref{eqn:covariant_b}.  Thus, the \CDF{} analysis
applies to our \Zprime{} boson only to the extent that initial-state
third-generation quarks do not contribute to dilepton production.
Since the top quark's parton distribution function is approximately
zero, we can reasonably apply the \CDF{} limits for
lepton/up-type-quark contact interactions to our model.  

Comparing the interactions \ref{eqn:LEQ} and \ref{eqn:ourzp}, we find the 
relationship between
$M_{\Zprime}$ and $\Lambda$ is
\begin{equation}
M_{\Zprime} = \Lambda \left(\frac{c_\phi}{s_\phi}\right) 
\sqrt{\frac{\alphaem}{{4 \sin^2\theta}}}
\, .
\label{eqn:zplimf}
\end{equation}
The \CDF{} bounds in Equation~\ref{eqn:cdflimit} imply
\begin{gather}
  M_{\Zprime} >\begin{cases} \left(\frac{c_\phi}{s_\phi}\right) \times
    \unit[380]{GeV} & \text{from dimuons}\\  \\
    \left(\frac{c_\phi}{s_\phi}\right)\times\unit[345]{GeV} &
    \text{from dielectrons}\ .\end{cases}
\end{gather}
These limits are comparable to those from the \LEP{} data for $s_\phi
\approx c_\phi$, but become significantly weaker at large $s_\phi$.

The \Dzero{} Collaboration has performed a similar analysis for high
energy dielectron production \cite{d0-dylim}, but assuming only
first-generation fermions participate in the contact interactions
(i.e.\ the terms explicitly written in Equation~\ref{eqn:LEQ}.  Since
they include only first-generation fermions, their limit
\begin{equation}
\Lambda(\elec_L; \qup_L, \qdown_L) > \unit[4.2]{TeV}
\label{eqn:dolimt}
\end{equation}
applies directly to our \Zprime{} boson, yielding the constraint
\begin{equation}
M_{\Zprime} > \left(\frac{c_\phi}{s_\phi}\right)\times\unit[390]{GeV}
\end{equation}
which is comparable to the result obtained by \CDF{}.

The \TEVtwoK{} Group Report \cite{tev2000} projects that the limits on
the scale of quark-lepton compositeness will be increased to $\Lambda
\geq \unit[6 - 7]{TeV}$.  This would raise the corresponding limits on
the mass of these \SUtwo{} \Zprime{} bosons to $M_{\Zprime} \geq
(c_\phi / s_\phi) \times \unit[550 - 650]{GeV} $.


\section{Direct Searches for an \SUtwo{} \Zprime{} in $
  \proton\Dbar{\proton} \longrightarrow \Zprime \longrightarrow
  \tauplus\tauminus$} 
\label{sec:channels} 

In studying direct production of a \Zprime{} boson from an extended
electroweak gauge structure, we must be aware of several competing
issues.  The couplings of third generation fermions to the extended
gauge sector are enhanced relative to their Standard Model values,
while those of the first and second generation particles are reduced.
Since the current lower bounds on the \Zprime{} mass are on the order
of $\unit[400]{GeV}$, the only machine presently available to perform
a direct search is the Fermilab Tevatron.  Thus, we are led to
searching for a clean signal in third generation final states in a
hadronic environment.  Given the high mass of the top quark, the large
QCD backgrounds for bottom production, and the difficulty of seeing
the \tauon{}-neutrino final state in photon plus missing energy or
monojet events, the most promising channel is
\begin{equation}
\proton\Dbar{\proton} \longrightarrow \Zprime \longrightarrow
\tauplus \tauminus + \particle{X}\ .
\end{equation}

Each \tauon{} decays to a final state including either hadrons or one
charged lepton and neutrinos.  Our analysis concentrates on fully
leptonic decays; we discuss possibilities with hadronic final states in
Section~\ref{sec:disc}.  The three fully leptonic final states
are characterized by opposite sign leptons with relatively large
missing energy:
\begin{gather}
\tautau \longrightarrow 
\begin{cases}
\muplus \muminus + \text{neutrinos}\\
\eplus \eminus + \text{neutrinos}\\
\eplus \muminus\ (\text{or } \eminus \muplus) + \text{neutrinos}\\
\end{cases}
\label{eqn:finalstates}
\end{gather}
Dimuon and dielectron final states are also characteristic of a number
of Standard Model processes (e.g., Drell-Yan), making it difficult to
separate our \Zprime{} signal from the backgrounds.  Thus, we 
focus on the last channel above, namely oppositely charged,
high-\pt{} electron-muon pairs.

We have employed \pythia{} version 6.127 \cite{pythia} with our own
simple model of the \Dzero{} detector to generate events and used our
own code to analyze the generated data.  Our idealized version of the
Run II \Dzero{} detector defines the fiducial volume and smears event
tracks.  The central calorimeters are taken to have a pseudorapidity
coverage, $\eta$, of $|\eta| \leq 1.1$, while the end-cap calorimeters
are taken to have a coverage of $1.5 \leq |\eta| \leq 4.0$ for jets
and $1.5 \leq |\eta| \leq 2.5$ for leptons.  The dilepton events may
initially be selected by a single or double lepton trigger.  Our
analysis assumes a trigger times offline selection efficiency of 95\%,
per lepton.  We choose to trigger on the transverse momentum of the
leptons, and set the triggering threshhold for each lepton at
$\unit[15]{GeV}$.\footnote{We have also considered a combination of
  this hard \pt\ trigger for one of the leptons and a softer \pt\ 
  trigger for the other.  While doing so does increase the number
  events to analyze, our later analysis cuts end up eliminating these
  extra events.}  Jet reconstruction efficiency is taken to be 100\%
for jets with transverse momenta in excess of
$\unit[8]{GeV}$.\footnote{We do not consider basing event triggers on
  jets, and so do not consider the jet triggering efficiency, which
  would be expected to be much lower than the efficiency for
  reconstruction given a previous trigger.  This is, however, an issue
  of some interest, and we will return to it below, in
  Section~\ref{sec:disc}.}  We identify a jet as a cluster of hadronic
energy in excess of $\unit[8]{GeV}$ contained within a cone of base
radius $R = \smash{\sqrt{\eta^2+\phi^2}}<1$. Our jet reconstruction
code is based on the \pythia{} cluster finding algorithm.  Based on
these assumptions, we chose as an event trigger the presence of an
electron and a muon of opposite electric charge both with transverse
momenta in excess of $\unit[15]{GeV}$, with tracks lying within the
fiducial volume of the detector.

We generated several sets of signal events corresponding to different
  \Zprime{} boson masses and different values of the mixing angle
  $\phi$ in the region least constrained by precision electroweak
  data (large $s_\phi$, see Figure~\ref{fig:ew-comp-limits}).  We also
  generated \elec\muon\ events from the four significant sources
  of Standard Model backgrounds
\begin{gather}
\proton\Dbar{\proton} \longrightarrow \Znaught/\gamma^* \longrightarrow \tautau
\longrightarrow \electron\muon \, + \, \text{neutrinos}\\
\proton\Dbar{\proton} \longrightarrow \Wplus\Wminus \longrightarrow
\electron\muon \, + \, \text{neutrinos}\\
\proton\Dbar{\proton} \longrightarrow \qtop\Dbar{\qtop}
\longrightarrow \Wplus\Wminus
\qbottom\Dbar{\qbottom}\longrightarrow \electron\muon
\qbottom\Dbar{\qbottom} \,+ \, \text{neutrinos}\\ 
\proton\Dbar{\proton} \longrightarrow \qbottom\Dbar{\qbottom}
\longrightarrow \electron\muon
\quark\Dbar{\quark} \,+ \, \text{neutrinos}\ ,
\end{gather}
which we will call the \Znaught, \Wpart-pair, top, and bottom backgrounds,
respectively.  The superficially similar backgrounds from charm
production were eliminated by the event selection cuts we discuss
below.  When generating prompt tau leptons from the signal and
background processes in Pythia, we have ignored polarization
correlations between the tau pairs.  This is a reasonable
approximation for our study, since the correlations are diluted when
considering only the leptonic decays of the tau.  A correct accounting
of the correlations will most likely improve the separation of signal
and background.

For each signal and background process, we generated a minimum of $3
\times 10^4$ events matching the $\electron\muon$ final state at
$\sqrt{s} = \unit[2]{TeV}$.  For each event we verified the presence
of the opposite charge $\electron\muon$ pair, and then reconstructed
the jets in the event.  To the four-momenta of these leptons and jets,
we applied smearing functions appropriate to the detector.\footnote{We
  have used the following algorithms for track smearing for our Run II
  detector: for electrons, we performed a Gaussian smearing based on
  the electron energy, with $\Delta E/E = 15\%/\sqrt{E}$; for muons,
  we performed a Gaussian smearing based on the transverse momentum,
  with standard deviation given by $\sigma_\ptmu = 1.5\times 10^{-3}
  \ptmu^2$, where the transverse momentum is measured in $\unit{GeV}$;
  finally, for jets we performed a Gaussian smearing based on the
  transverse jet energy, with standard deviation quadratic in the
  transverse energy with $\eta$ dependent coefficients.}  We accepted
events in which the smeared tracks of both the electron and the muon
lay in the fiducial volume of our idealized Run II \Dzero{} detector.
Smeared jets falling outside the detector were dropped from the
events.  We eliminated events in which both smeared leptons no longer
passed the trigger cuts.  Finally, the four momenta of both leptons
and the surviving jets in the remaining events were stored for later
``offline'' analysis.

\begin{figure}[tb]
\begin{center}
  \begin{minipage}{(\textwidth-1in)/3}
    \begin{center}
      \includegraphics[width=\textwidth]{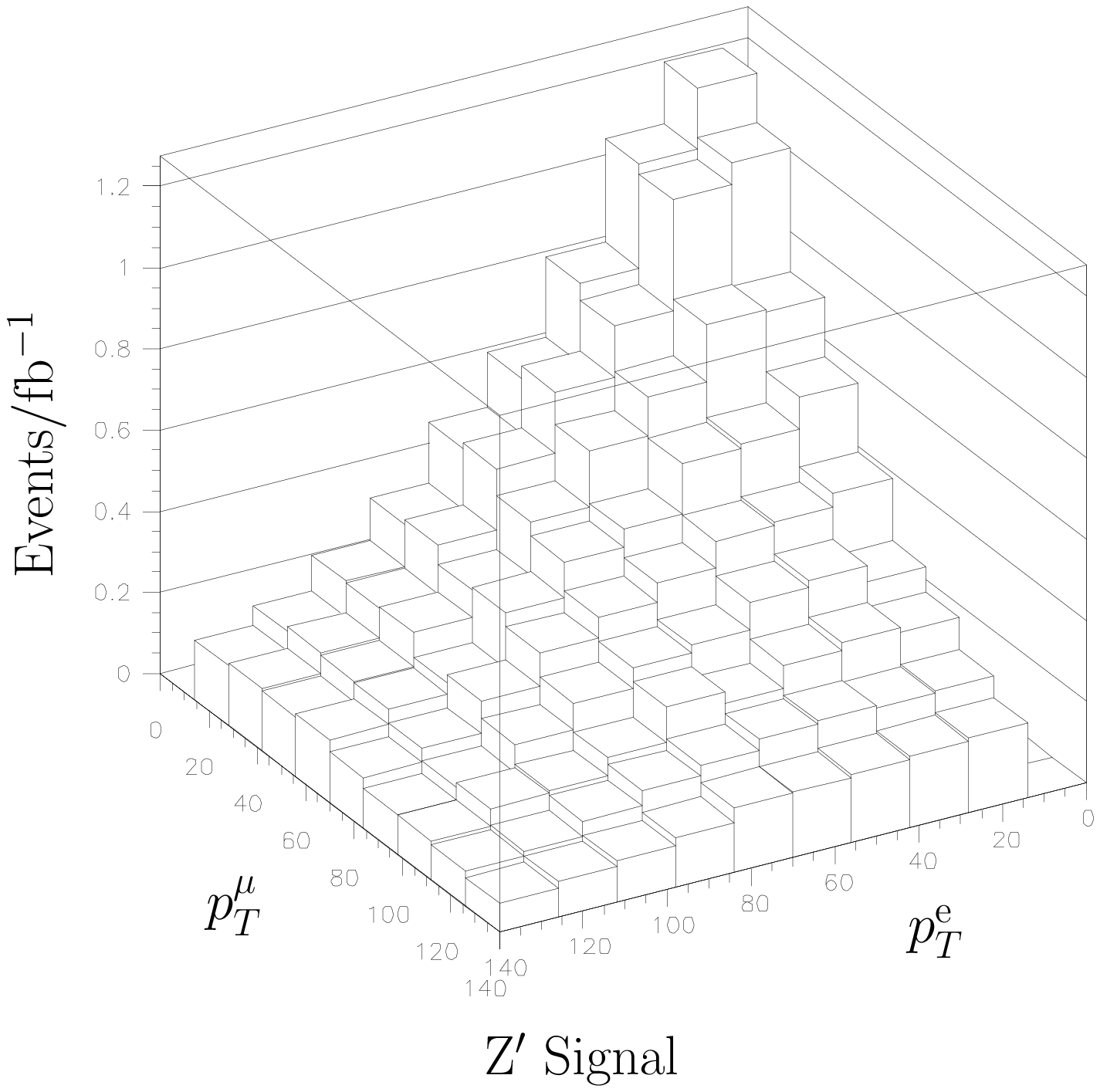}
    \end{center}
  \end{minipage}\qquad
  \begin{minipage}{(\textwidth-1in)/3}
    \begin{center}
      \includegraphics[width=\textwidth]{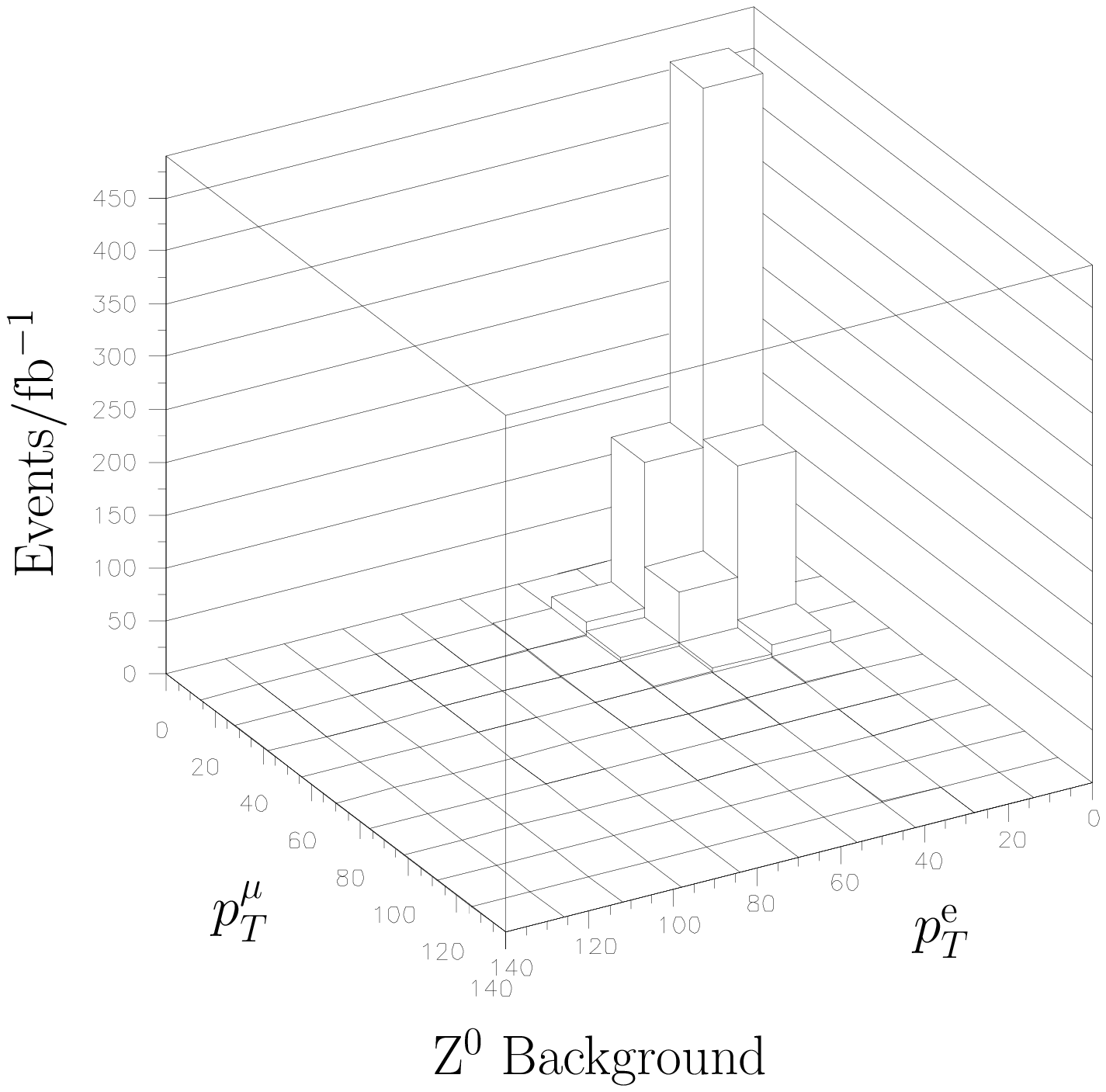}
    \end{center}
  \end{minipage}\qquad
  \begin{minipage}{(\textwidth-1in)/3}
    \begin{center}
      \includegraphics[width=\textwidth]{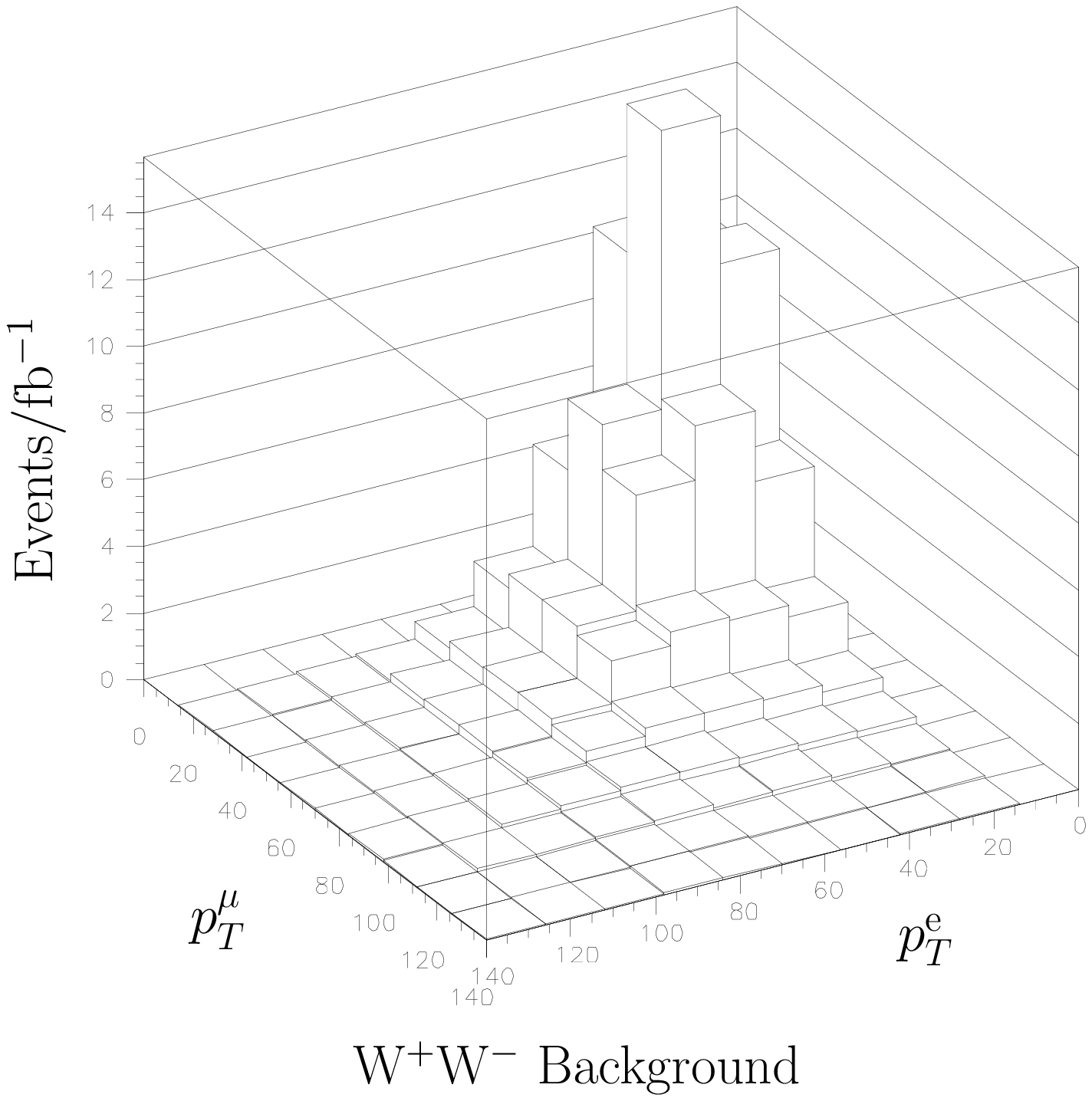}
    \end{center}
  \end{minipage}\\
  \begin{minipage}{(\textwidth-1in)/3}
    \begin{center}
      \includegraphics[width=\textwidth]{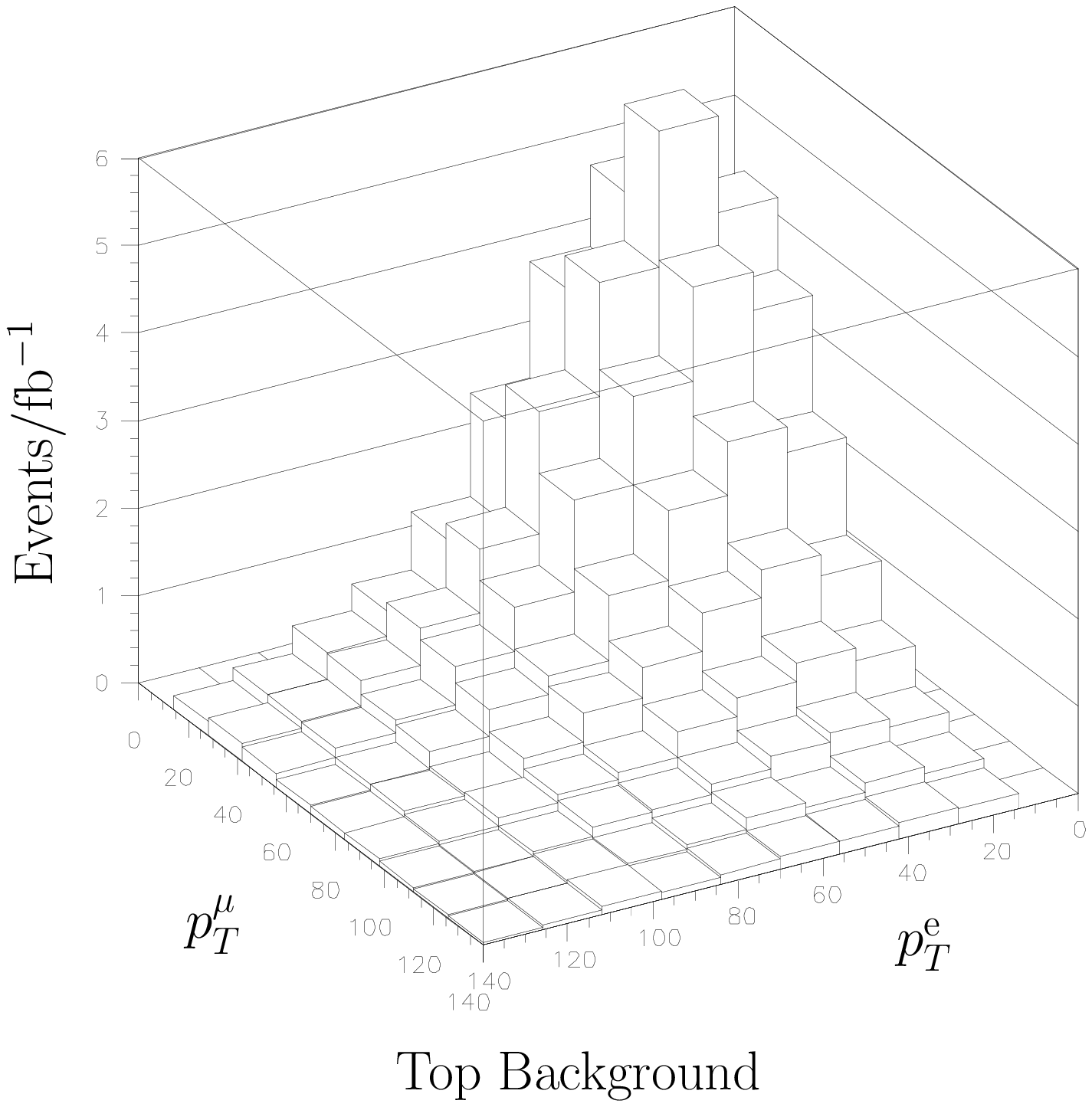}
    \end{center}
  \end{minipage}\qquad
  \begin{minipage}{(\textwidth-1in)/3}
    \begin{center}
      \includegraphics[width=\textwidth]{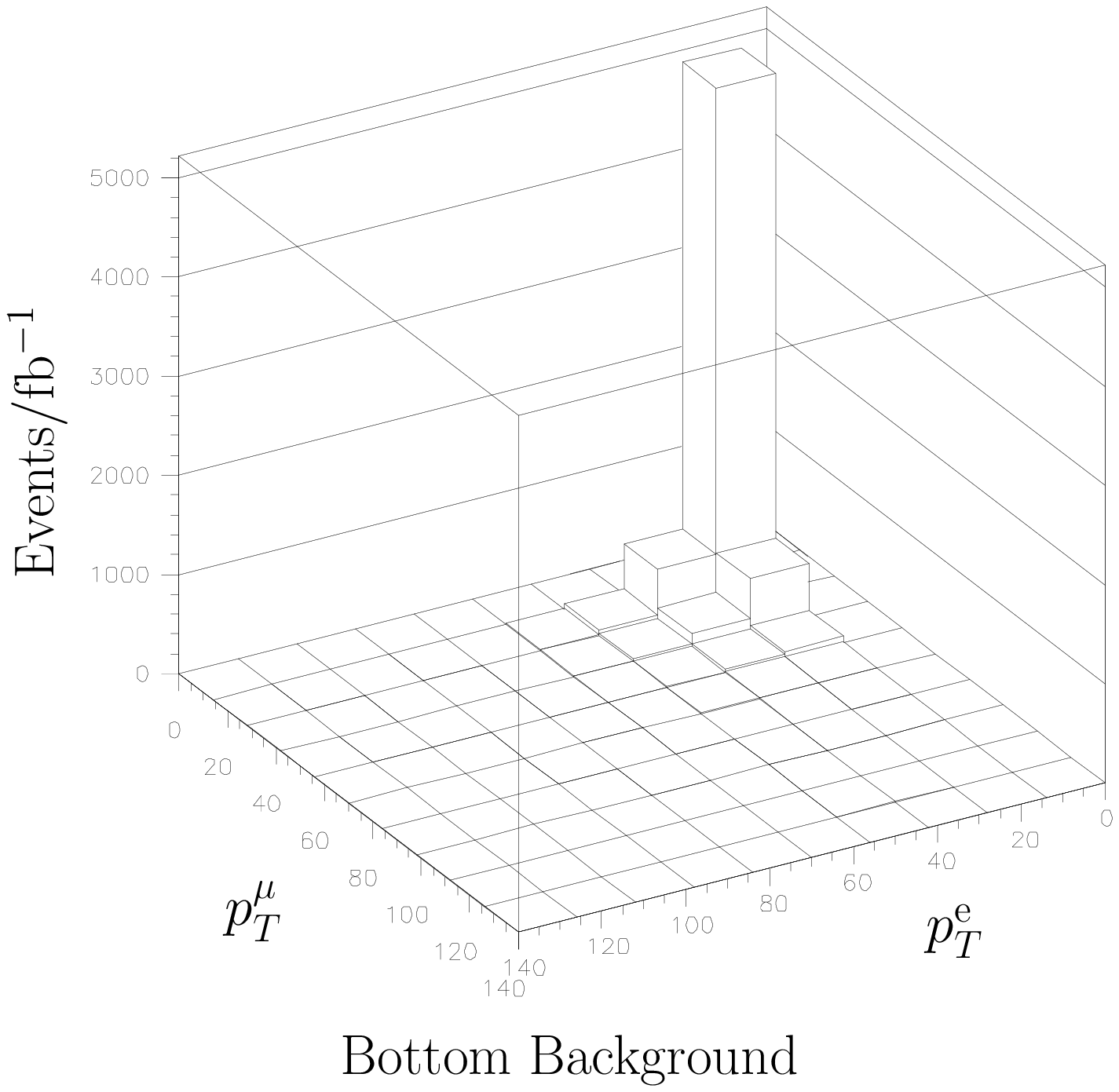}
    \end{center}
  \end{minipage}
\end{center}
\caption{LEGO plot of the distribution of events in the
  \pte-\ptmu{} plane.  The vertical axis displays the cross section
  ($\text{Events}/\unit{fb^{-1}}$), the left horizontal axis the muon
  transverse momentum (\ptmu{} in $\unit{GeV}$), and the right
  horizontal axis the electron transverse momentum (\pte{} in
  $\unit{GeV}$).  Both horizontal axes run from \unit[0]{GeV} up to
  \unit[140]{GeV}.  The upper left plot is for the \Zprime{} signal
  process, upper center the \Znaught{} background, upper right the
  \Wpart-pair background, lower left the top background, and lower right
  the bottom background.  Notice the symmetry in the radial (\ptl)
  direction.  A cut on \ptl{} will eliminate the majority of the
  \Znaught{} and bottom backgrounds and a substantial part of the
  \Wpart-pair background.  The \Zprime\ plot is for $M_\Zprime =
  \unit[450]{GeV}$ and $s_\phi = 0.80$.}
\label{fig:trigger}
\end{figure}

\begin{figure}[tb]
\begin{center}
\includegraphics[width=\textwidth-2in]{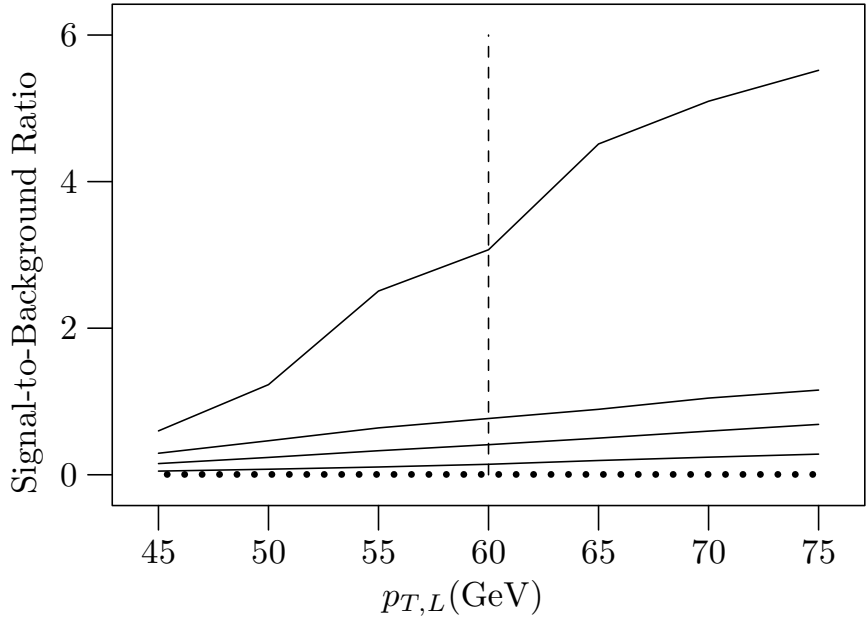}
\end{center}
\caption{Signal-to-background ratios as a function of the \ptl\ cut.
  The five curves correspond, from bottom to top, to applying
  successive analysis cuts.  The lowest (dotted) curve shows the data
  following acceptance and trigger cuts only.  The second curve is
  obtained after applying the \ptl\ cut specified on the $x$-axis.
  The third curve is obtained after applying the jet multiplicity cut,
  $n_{\rm jets}<2$, to the events surviving the \ptl\ cut.  The fourth
  curve is obtained after applying the topological cut on the
  \elec\muon\ opening angle, $\cos\theta_{\elec\muon} < -0.5$, to the
  events surviving the \ptl\ and jet multiplicity cuts.  The topmost
  curve results from applying the cut on the opening angle between the
  low energy lepton and the missing transverse energy in the event,
  $\cos\theta_{\ell \not{E}_T} > 0.9$, to the events surviving the
  \ptl\, jet multiplicity, and \elec\muon\ opening angle cuts.  The
  choice of \pthresh{} was made by maximizing the signal-to-background
  ratio while maintaining sufficient signal count for a large range of
  Z' masses.  We chose $\pthresh = \unit[60]{GeV}$, which is indicated
  on the plot by the dashed vertical line.  This plot is for
  $M_{\Zprime} = \unit[450]{GeV}$ and $s_\phi = 0.80$.}
\label{fig:soverb}
\end{figure}

Properly normalized \pt{} distributions of background and signal
events following the trigger stage are shown in
Figure~\ref{fig:trigger}.
Examination of the smeared trigger distributions in this figure
suggests offline analysis cuts that will eliminate the majority of the
pure electroweak and \qbottom\ backgrounds, while preserving
sufficient signal to permit analysis.  These \pt{} distributions are
symmetric in the ``radial'' direction in the \pte-\ptmu{} plane, where
\pte{} (\ptmu) is the transverse momentum of the electron (muon).  For
our primary analysis cut, we define the leptonic transverse momentum,
\ptl, of the event,
\begin{equation}
\ptl = \sqrt{\pte^2 + \ptmu^2}\ ,
\label{eqn:ptl}
\end{equation}
where we require \ptl{} to exceed some threshold, \pthresh.  Our
choice of \pthresh{} was dictated by the requirements of enhancing the
signal-to-background ratio while maintaining a sufficiently high
absolute signal event rate, and was chosen in conjunction with the
other cuts to be described below.  We display typical
signal-to-background rates before and after the \ptl{} cut in
Figure~\ref{fig:soverb}.
Based on the calculated signal-to-background ratio and the absolute
signal rates, we placed our \ptl{} cut at
\begin{equation}
\ptl \ge \unit[60]{GeV}\ .
\end{equation}
This value for \pthresh\ should effectively reduce the \Wplus\Wminus\
background, since some of the leptons from \Wpart-pair decay exhibit
a characteristic Jacobian peak near $M_W/2$.

\begin{figure}[tb]
\begin{center}
  \begin{minipage}{(\textwidth-1in)/3}
    \begin{center}
      \includegraphics[width=\textwidth]{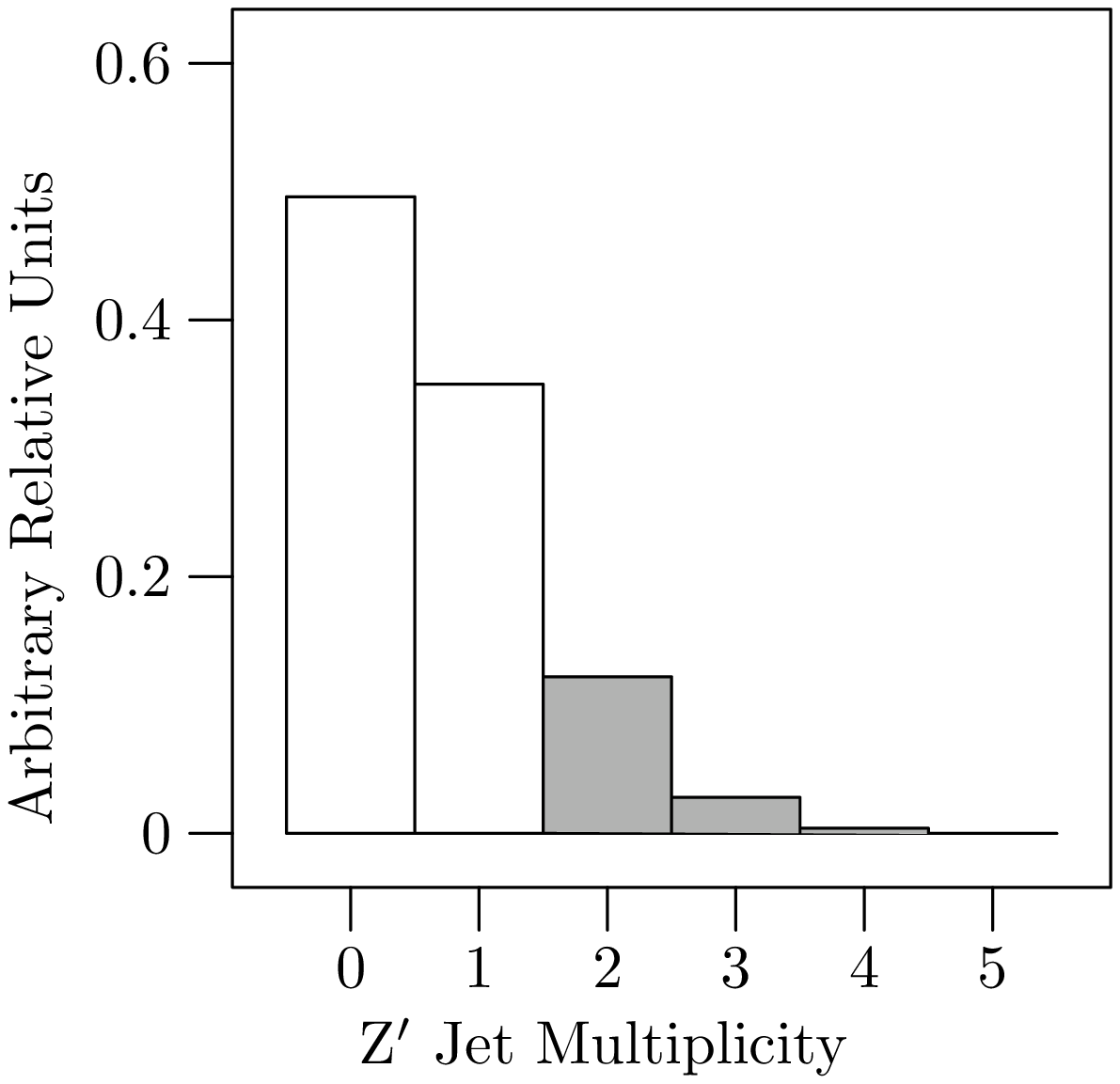}
    \end{center}
  \end{minipage}\qquad
  \begin{minipage}{(\textwidth-1in)/3}
    \begin{center}
      \includegraphics[width=\textwidth]{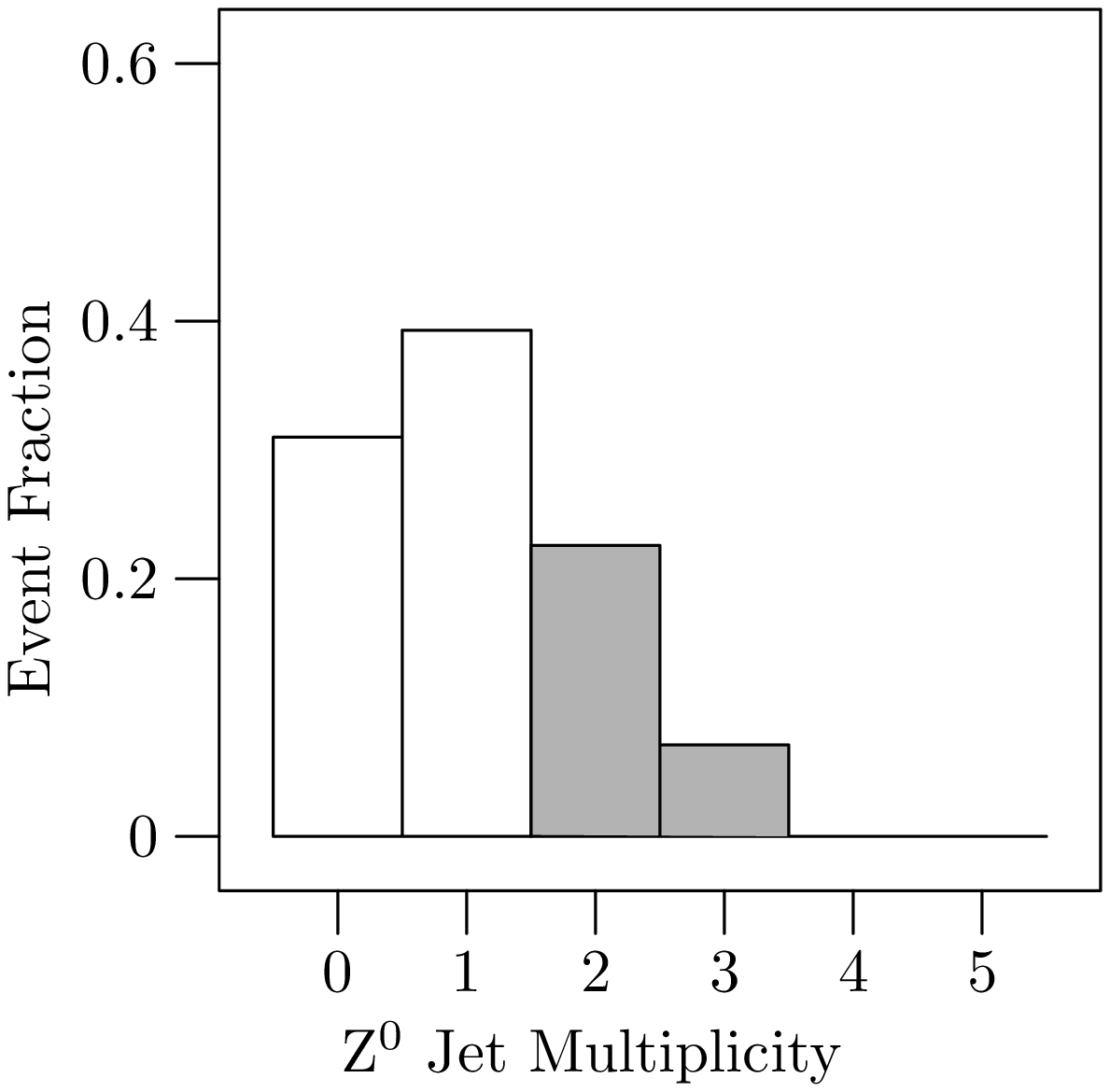}
    \end{center}
  \end{minipage}\qquad
  \begin{minipage}{(\textwidth-1in)/3}
    \begin{center}
      \includegraphics[width=\textwidth]{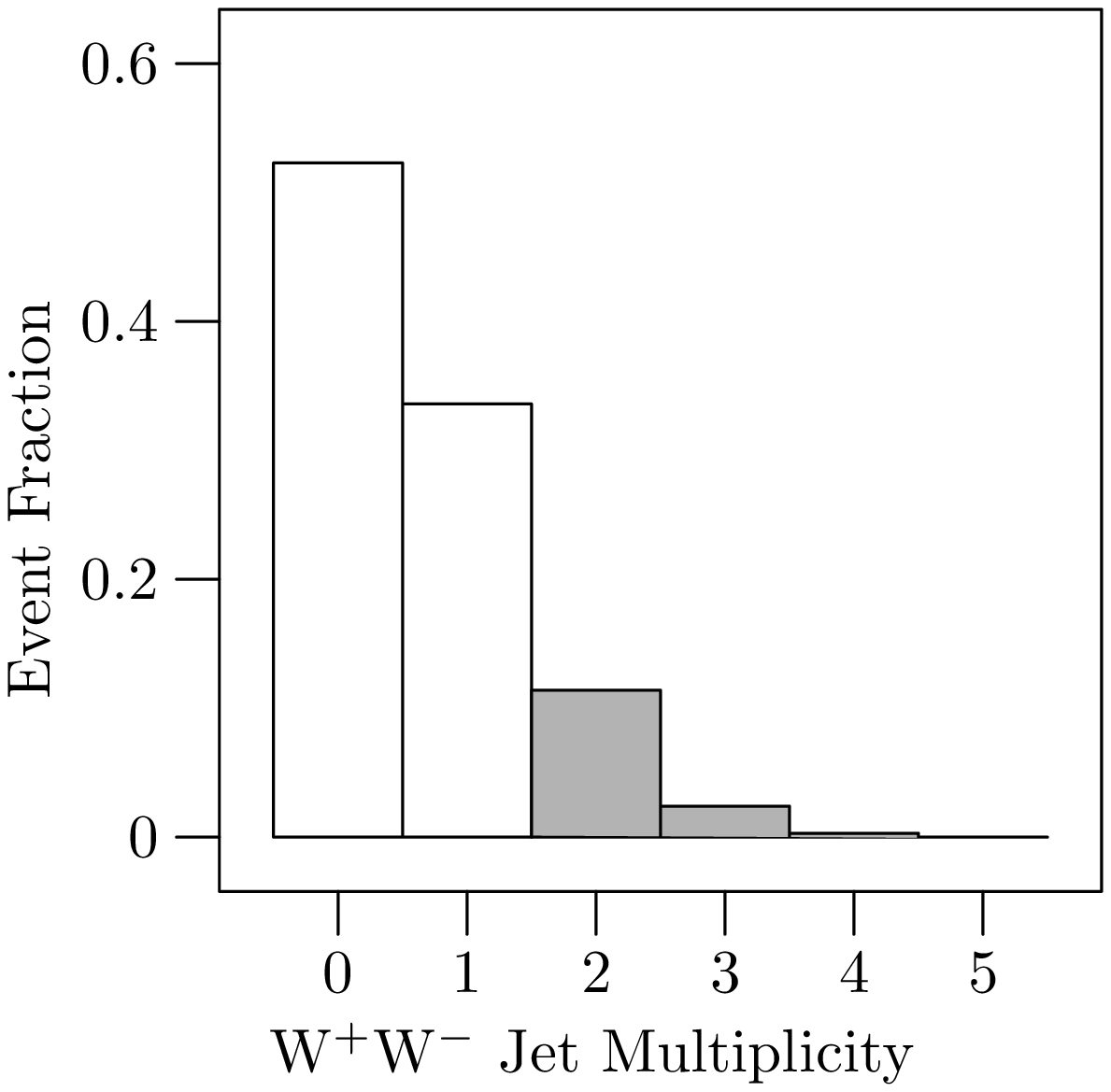}
    \end{center}
  \end{minipage}\\
  \begin{minipage}{(\textwidth-1in)/3}
    \begin{center}
      \includegraphics[width=\textwidth]{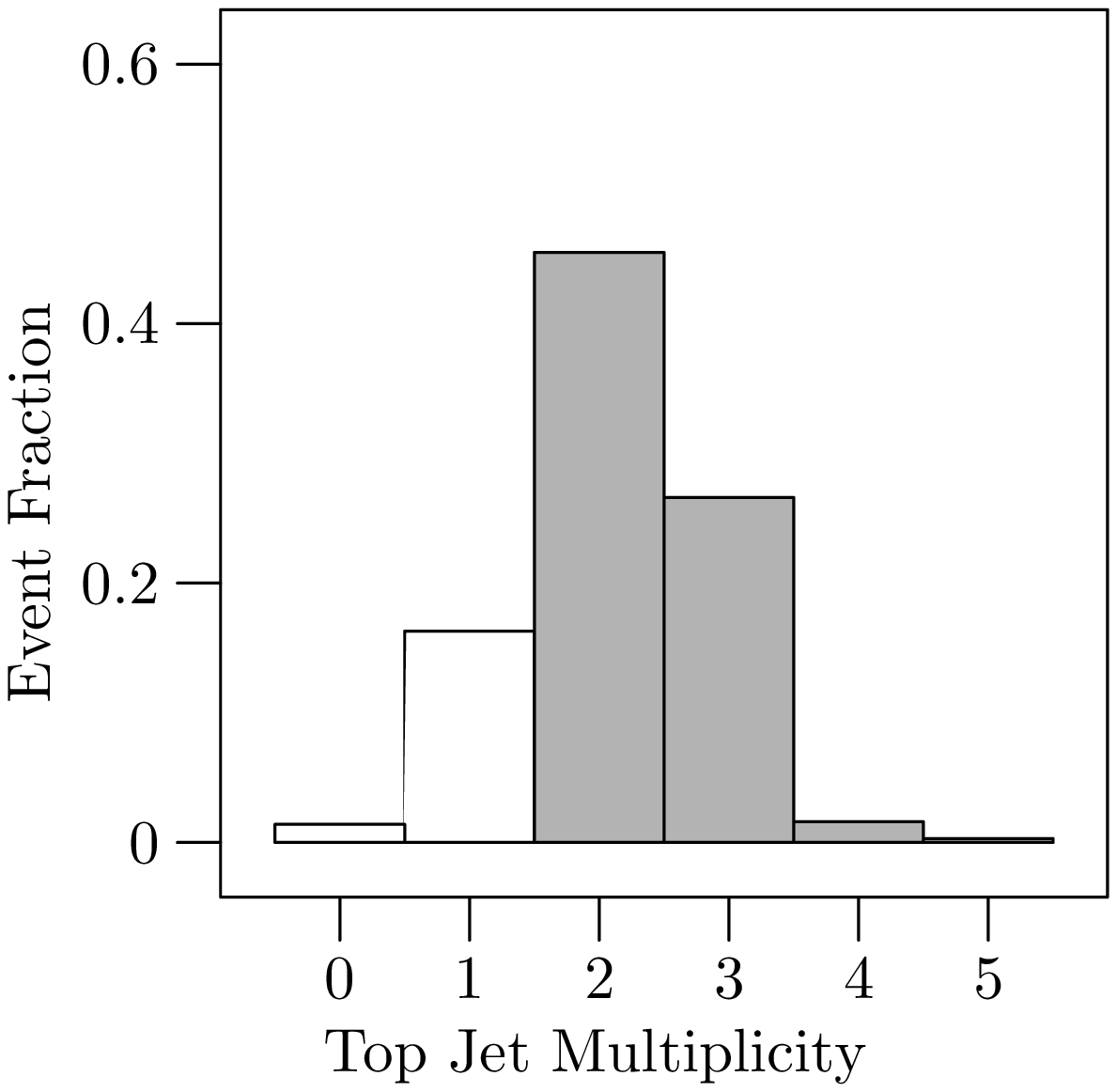}
    \end{center}
  \end{minipage}\qquad
  \begin{minipage}{(\textwidth-1in)/3}
    \begin{center}
      \includegraphics[width=\textwidth]{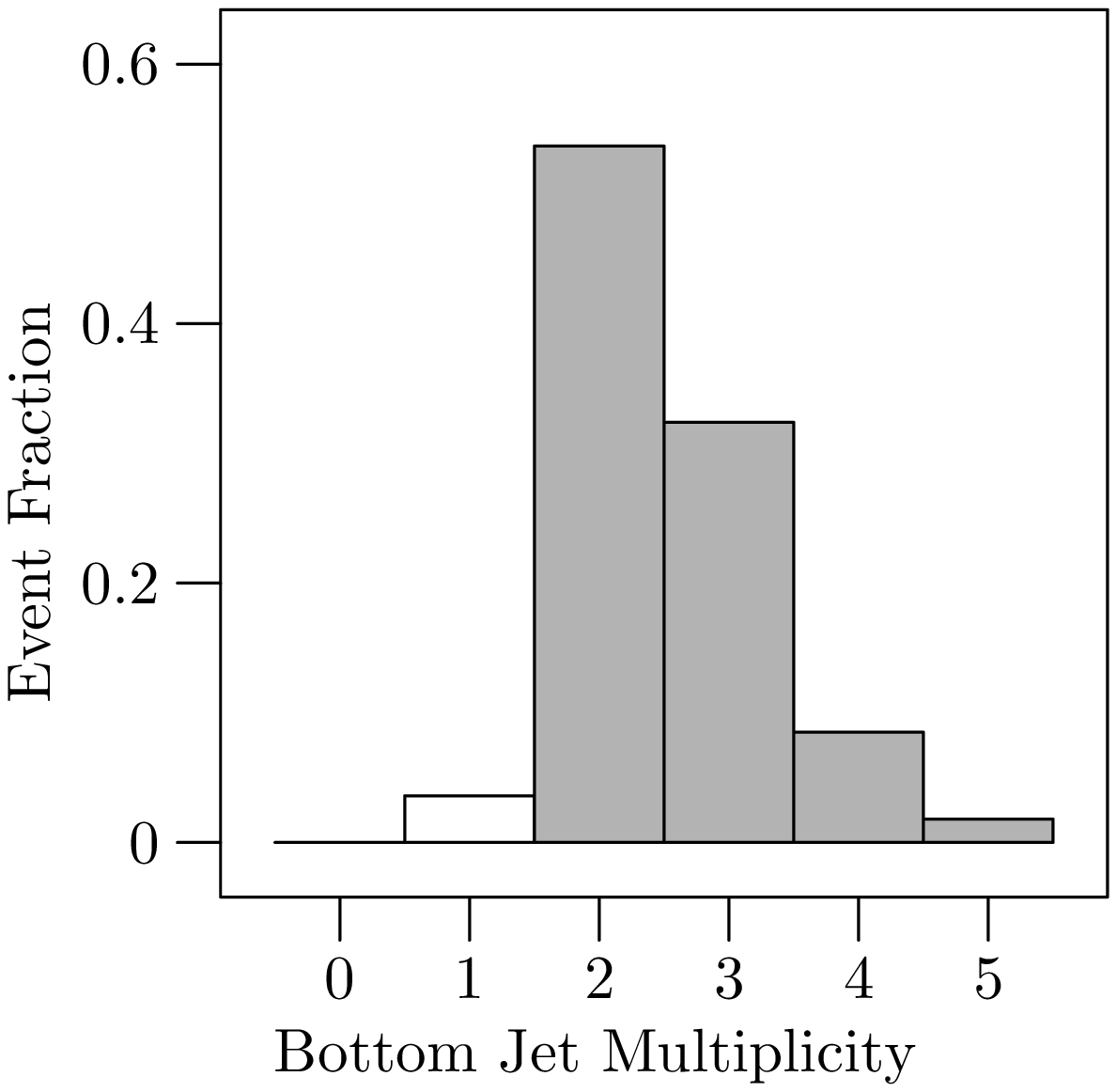}
    \end{center}
  \end{minipage}
\end{center}
\caption{Jet multiplicity for the signal and for each major Standard
  Model background process following the \ptl\ cut.  The upper left
  plot is for the \Zprime{} signal process, upper center the
  \Znaught{} background, upper right the \Wpart-pair background, and
  lower left the top background, and the lower right the bottom
  background.  As discussed in the text, all jets of energy exceeding
  $\unit[8]{GeV}$ are included.  Each bin is plotted by the fraction
  of events it contains; the final bin contains all events with $5$ or
  more jets.  Note the qualitative difference between the top and
  bottom distributions and the other three distributions, which
  suggests that events with more than one jet be eliminated (these are
  shaded in gray).  The \Zprime\ plot is for $M_\Zprime =
  \unit[450]{GeV}$ and $s_\phi = 0.80$.}
\label{fig:jets}
\end{figure}

To further improve signal purity, we consider the presence of hadronic
jet activity.  For our signal events, we would expect to see no
hadronic jet activity originating at the partonic event level.
Similarly, we expect no jet activity for the \Znaught{} and
\Wpart-pair backgrounds.  However, we always expect activity
associated with the top and bottom backgrounds.  In particular, we
expect two \qbottom{} jets associated with the top decays to \Wpart-b,
and two \qcharm\ jets associated with the bottom decays.  N\"aively
then, a cut on jet multiplicity will preferentially remove the top and
bottom backgrounds.  We have analyzed the expected jet distributions
of events surviving the \ptl\ cut, as measured in our simulation for
each type of event considered.  This includes the extra jet activity
generated by parton showering.  We display these distributions in
Figure~\ref{fig:jets}; again, by jets we mean here clusters of
hadronic activity with energy in excess of $\unit[8]{GeV}$.

Note that by rejecting events with jet multiplicity greater than one,
we can remove a large majority of the top and bottom backgrounds while
minimally impacting the strength of the signal.  Comparing the
signal-to-background ratio for our model before and after a jet
multiplicity cut, we find a significant increase in signal purity,
Figure~\ref{fig:soverb}.\footnote{In
  our simulations we do not account for effects due to pileup and
  multiple interactions on jet reconstruction.  we expect these issues
  will have minor impact on the final efficiency and the
  signal-to-background ratio.}

\begin{figure}[tb]
\begin{center}
  \begin{minipage}{(\textwidth-1in)/3}
    \begin{center}
      \includegraphics[width=\textwidth]{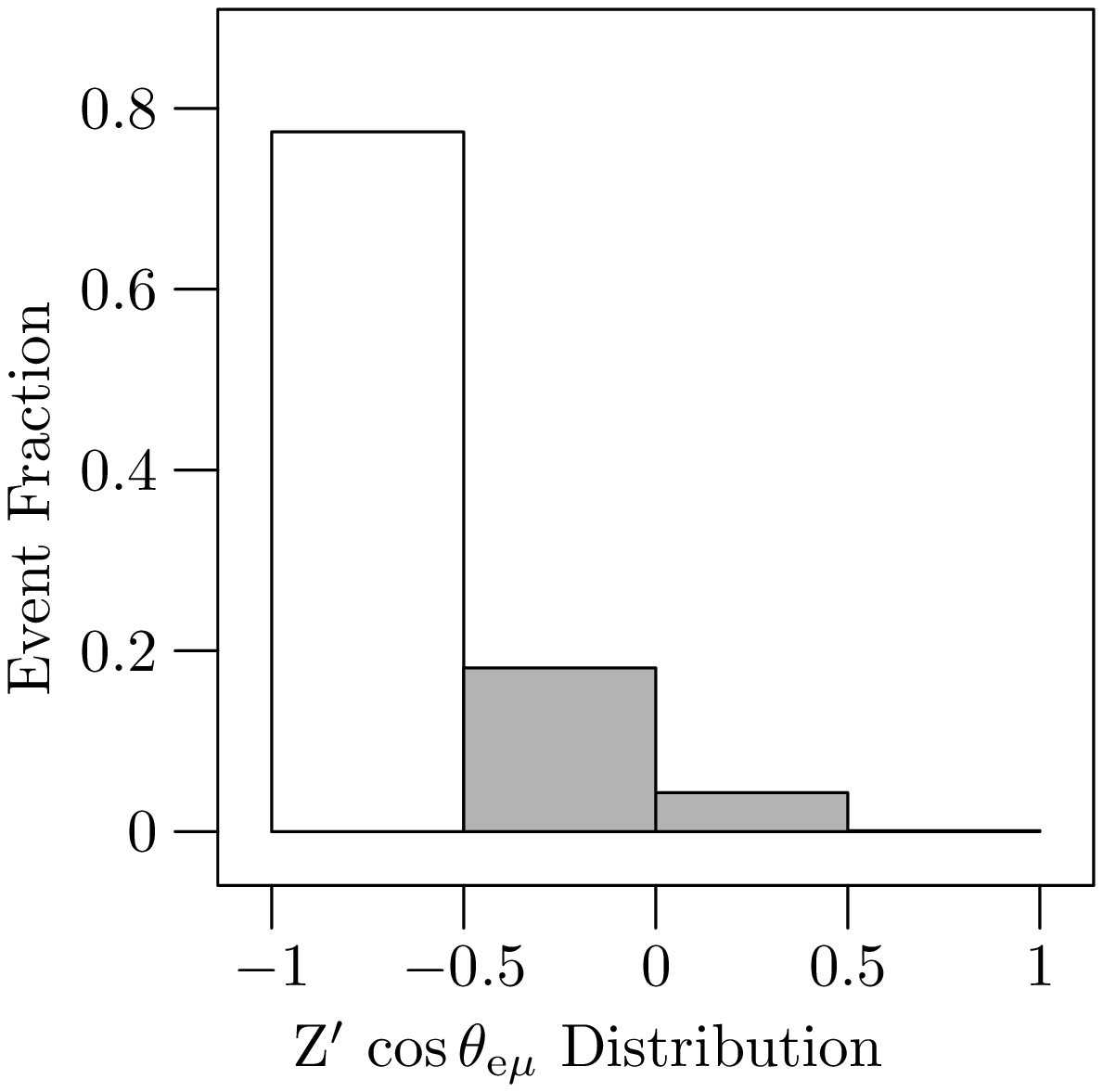}
    \end{center}
  \end{minipage}\qquad
  \begin{minipage}{(\textwidth-1in)/3}
    \begin{center}
      \includegraphics[width=\textwidth]{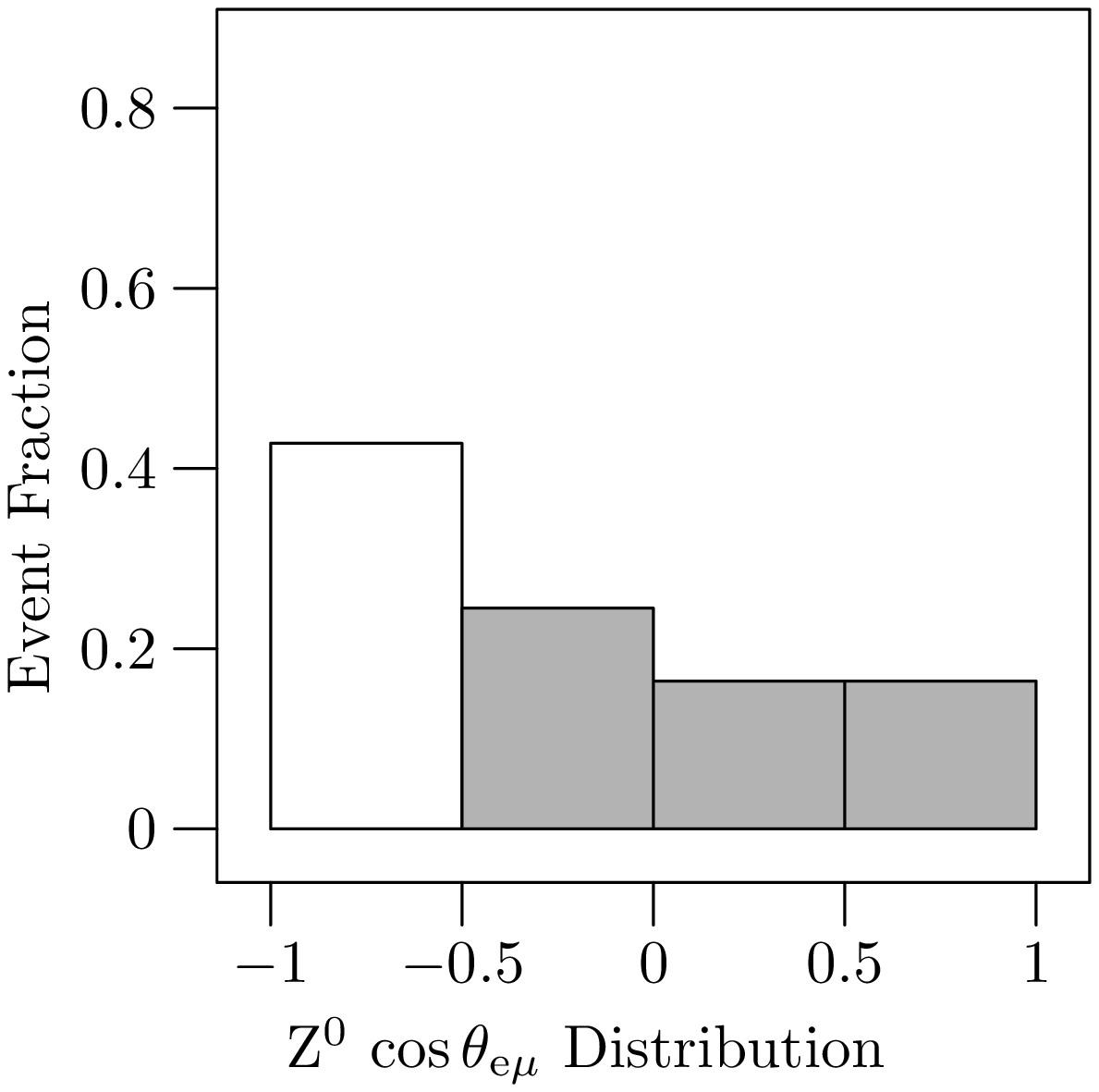}
    \end{center}
  \end{minipage}\qquad
  \begin{minipage}{(\textwidth-1in)/3}
    \begin{center}
      \includegraphics[width=\textwidth]{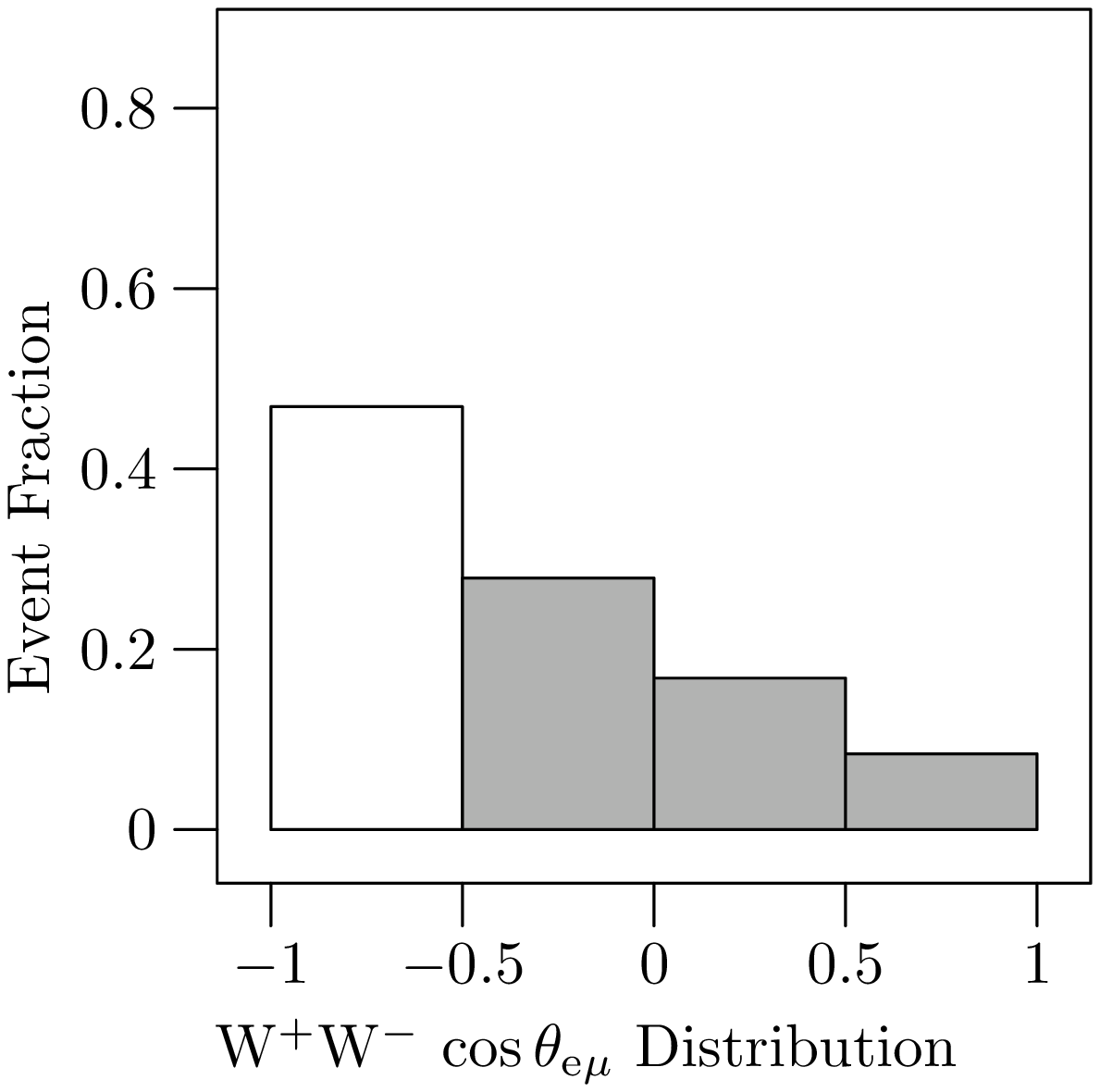}
    \end{center}
  \end{minipage}\\
  \begin{minipage}{(\textwidth-1in)/3}
    \begin{center}
      \includegraphics[width=\textwidth]{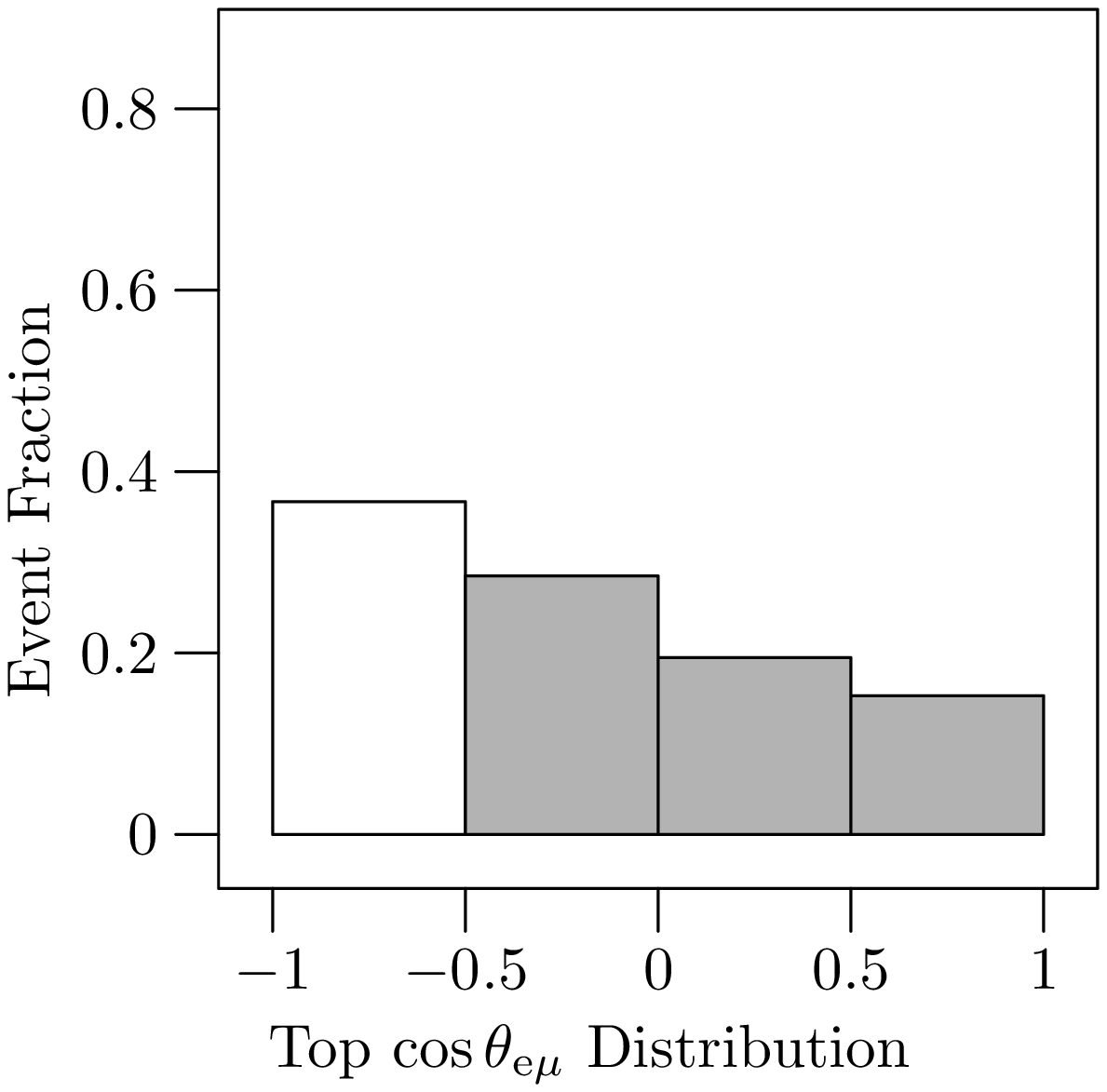}
    \end{center}
  \end{minipage}\qquad
  \begin{minipage}{(\textwidth-1in)/3}
    \begin{center}
      \includegraphics[width=\textwidth]{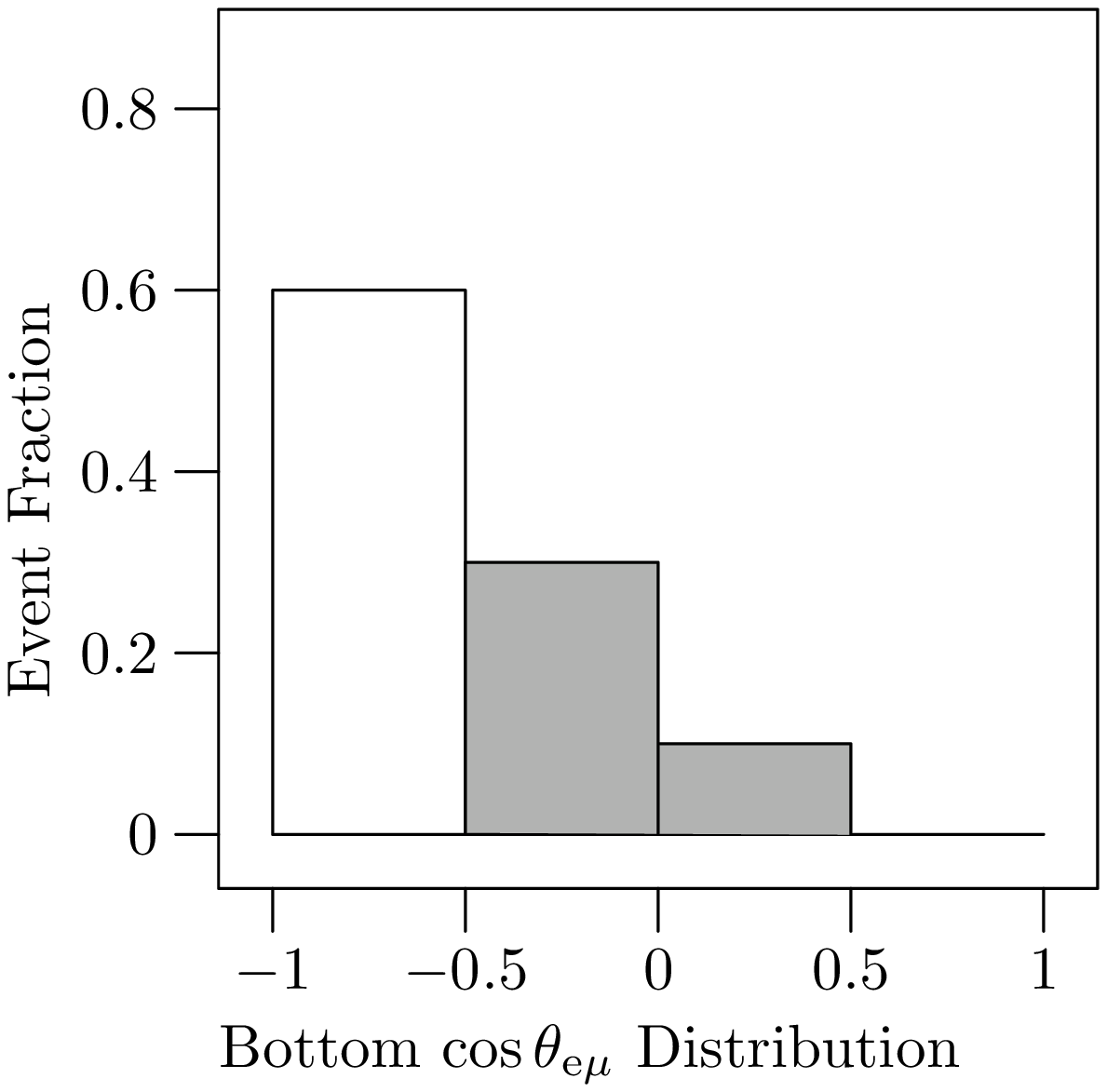}
    \end{center}
  \end{minipage}
\end{center}
\caption{Distribution of opening angles between the electron and muon
  tracks.  We have plotted here the cosines of the opening angle,
  $\cos\theta_{\elec\muon}$, split into two backward (negative
  cosine), and two forward (positive cosine) bins, following both
  the \ptl\ and jet multiplicity cuts.  The upper left plot is the
  \Zprime{} signal process, upper center the \Znaught{} background,
  upper right the \Wpart-pair background, and lower left the top
  background, and the lower right the bottom background.  Note that
  the vast majority, 80\%, of \elec\muon\ pairs from \Zprime\ signal
  events are in the ``strongly backward'' bin (with $\cos\theta <
  -0.5$), while approximately 40\% of \elec\muon\ pairs for all
  types of background events (except bottom, 60\%) are in the
  ``strongly backward'' bin.  A cut requiring the event to lie in the
  ``strongly backward'' bin will eliminate a large group of
  remaining background events (this group is shaded gray in each
  figure).  The \Zprime\ plot is for $M_\Zprime = \unit[450]{GeV}$ and
  $s_\phi = 0.80$.}
\label{fig:emuangle}
\end{figure}

Next, we apply a topological cut based on the opening angle between
the electron and muon, which we will label $\theta_{\elec\muon}$.
Given the large mass of the \Zprime{}, we expect it to be produced
nearly at rest in the detector, and expect the tau pair to be produced
back-to-back and highly boosted.  Because of this boost, the electron
and muon should be travelling nearly collinearly with their respective
parent taus, making them approximately back-to-back with one another
in the signal events.  The background events would not be expected to
have an opening angle distribution that is so highly peaked
back-to-back.  This is borne out by the Monte Carlo simulation.  In
Figure~\ref{fig:emuangle}, we display the distribution of opening
angles angles for events which have already passed the \ptl\ and jet
multiplicity cuts.  We choose to eliminate those events with
$\cos\theta_{\elec\muon} > -0.5$, that is, those where the electron
and muon are not strongly back-to-back.  Note that the vast
majority of the signal events pass the topological cut, while the
background events are more likely to fail it.
We have displayed the impact of this topological cut on the
signal-to-background ratios for various \ptl\ cuts in
Figure~\ref{fig:soverb}.

\begin{figure}
\begin{center}
  \begin{minipage}{(\textwidth-1in)/3}
    \begin{center}
      \includegraphics[width=\textwidth]{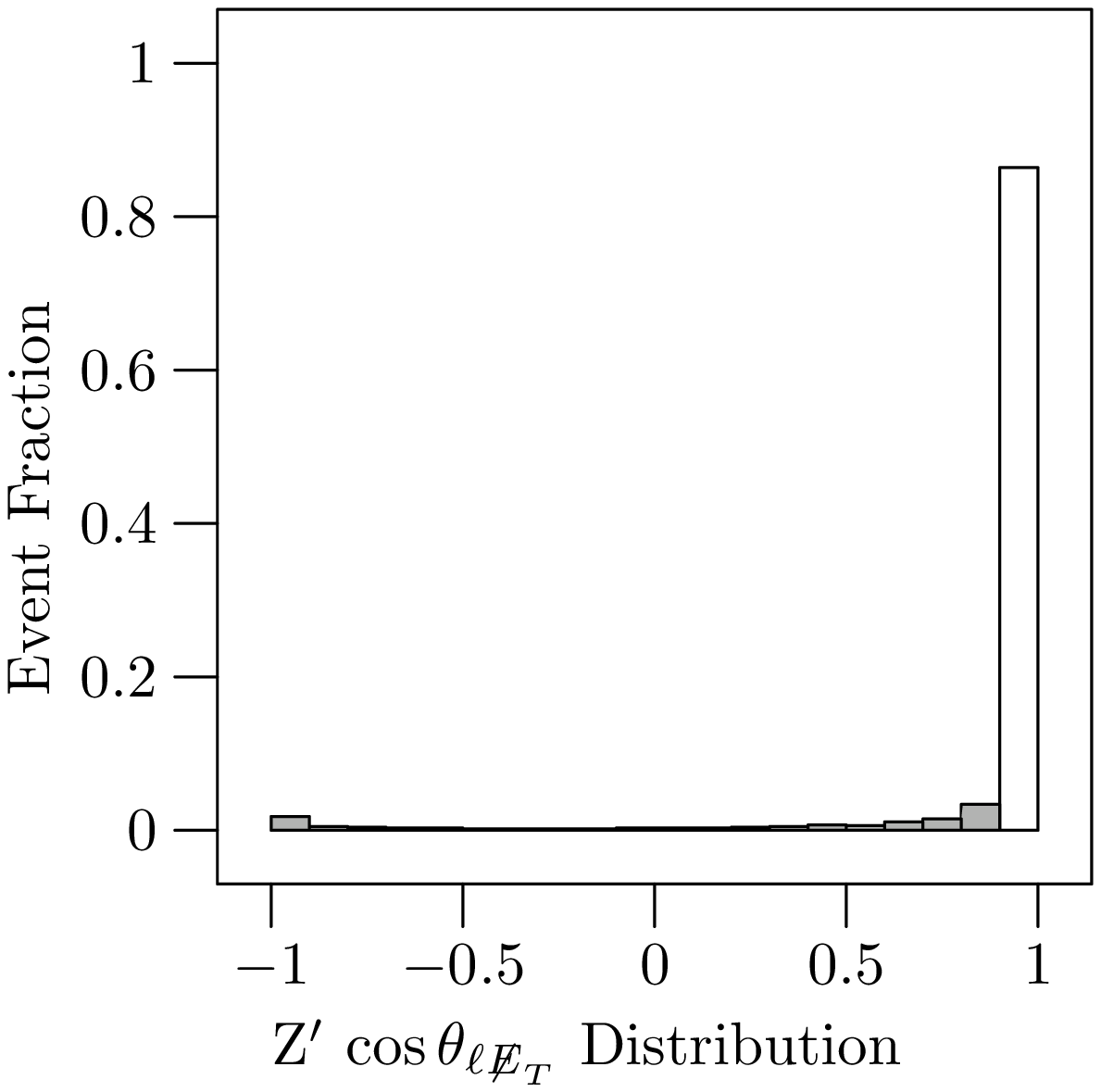}
    \end{center}
  \end{minipage}\qquad
  \begin{minipage}{(\textwidth-1in)/3}
    \begin{center}
      \includegraphics[width=\textwidth]{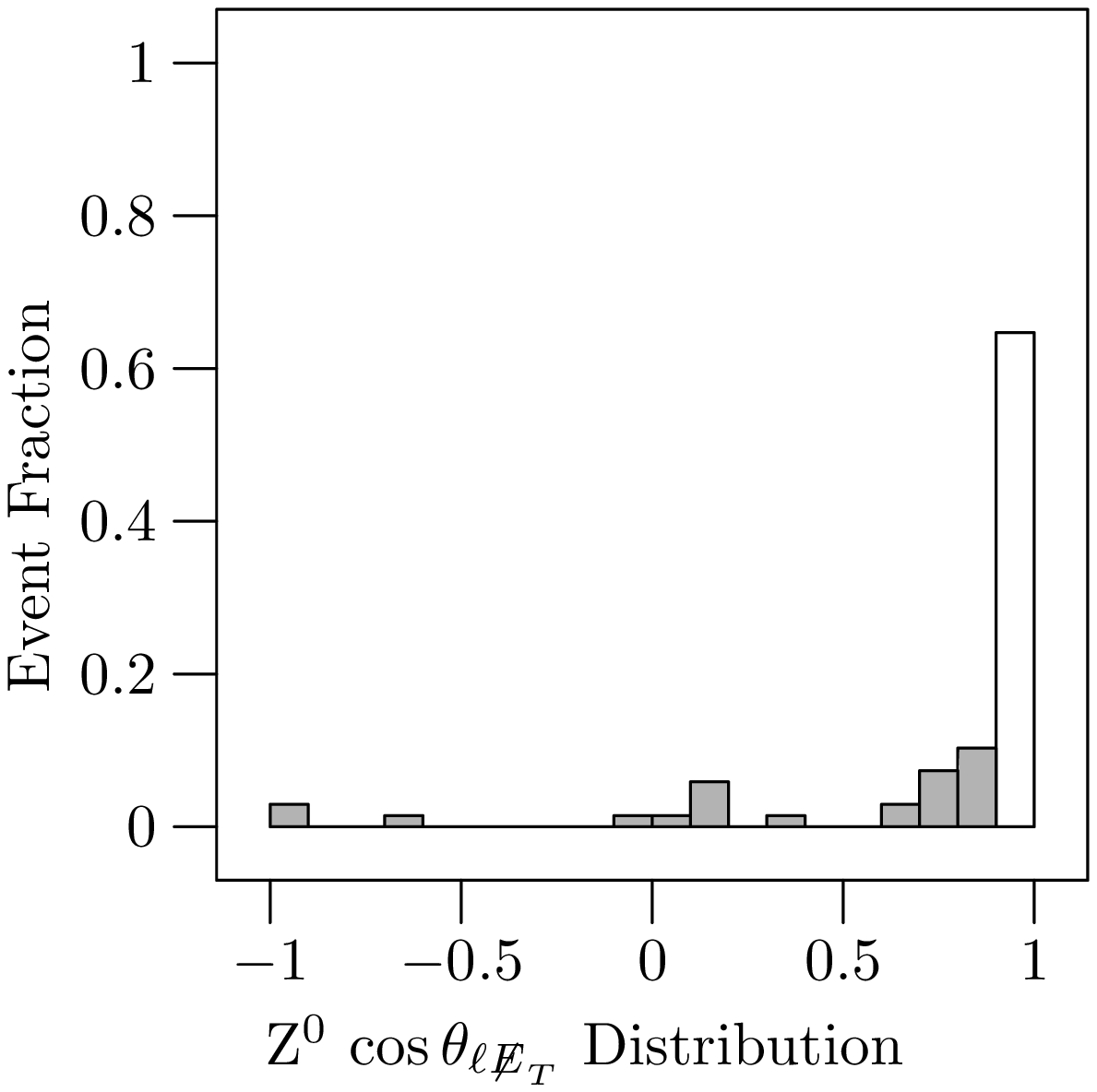}
    \end{center}
  \end{minipage}\\
  \begin{minipage}{(\textwidth-1in)/3}
    \begin{center}
      \includegraphics[width=\textwidth]{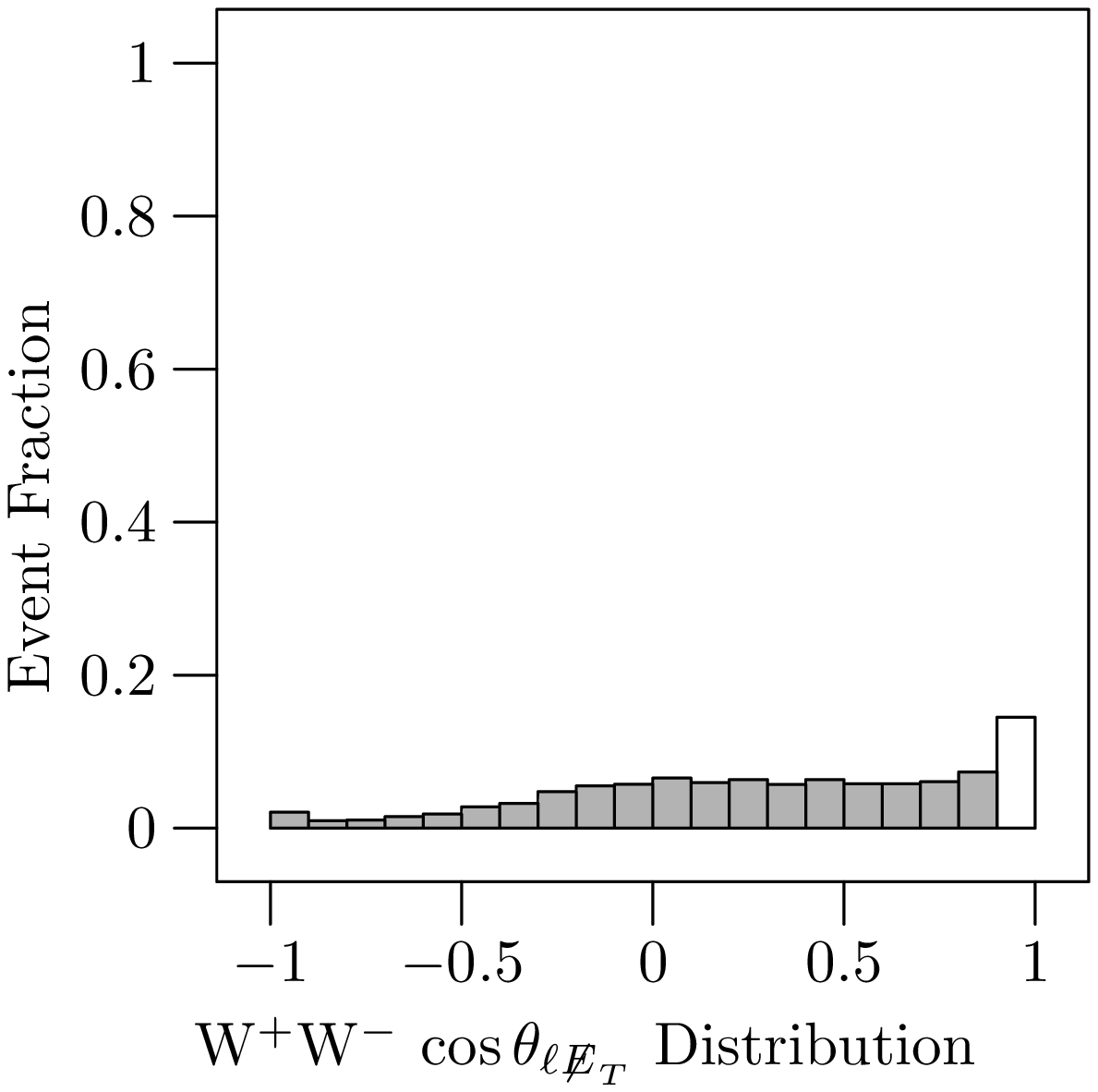}
    \end{center}
  \end{minipage}\qquad
  \begin{minipage}{(\textwidth-1in)/3}
    \begin{center}
      \includegraphics[width=\textwidth]{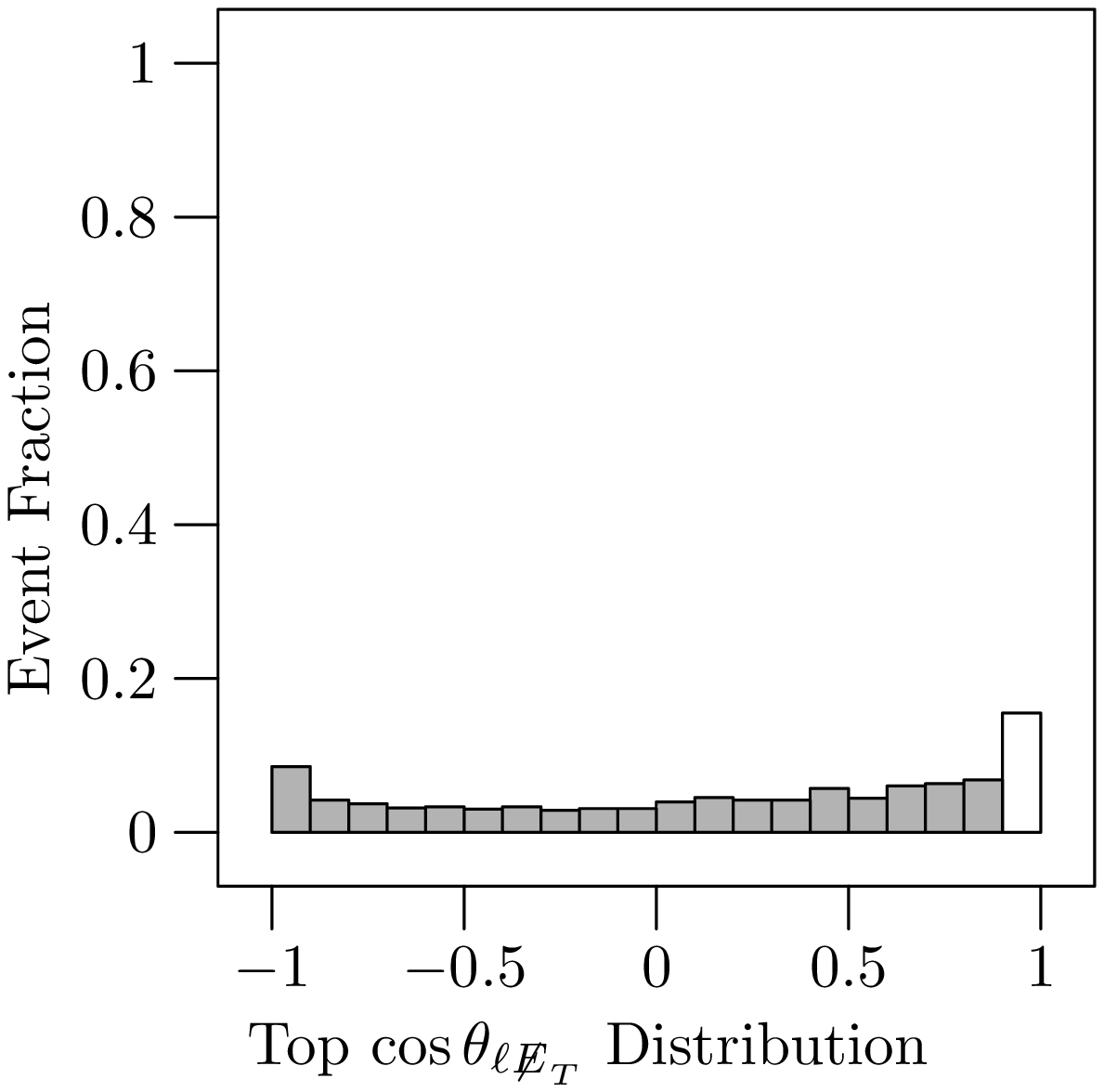}
    \end{center}
  \end{minipage}
\end{center}
\caption{Distribution of $\cos\theta_{\ell\mett}$, the opening angle
  between the lowest energy lepton track and the missing energy
  direction.  We have plotted here the fraction of events with given
  cosine of the opening angle, split into twenty equal bins, following
  the \ptl, jet multiplicity, and $\theta_{\elec\muon}$ cuts.  The
  upper left plot is the \Zprime\ signal process, upper right the
  \Znaught\ background, lower left the \Wpart-pair background, and lower
  right the top background.  The bottom background is also reduced by
  this cut (see Table~\ref{table:efficiency}), but the bottom
  background Monte Carlo sample did not have enough events remaining
  to produce a smooth distribution at the 20 bin granularity, and so
  is not plotted here.  As discussed in the text, we find the expected
  qualitative difference between the signal events and the now
  dominant \Wpart-pair background.  A cut at small opening angle will
  preferentially eliminate the \Wpart-pair background, and so we choose to
  require $\cos\theta_{\ell\mett}>0.90$.  The \Zprime\ plot is for
  $M_\Zprime = \unit[450]{GeV}$ and $s_\phi = 0.80$.}
\label{fig:lepmetangle}
\end{figure}

At this point, the remaining background is almost purely \Wpart-pair.  We
apply a final topological that that eliminates much of the remaining
\Wpart-pair background, based on the opening angle between the lowest \pt\ 
lepton and the transverse missing energy in the event, which we label
$\theta_{\ell\mett}$.  In order for the event to conserve momentum
overall, we would expect the missing transverse energy vector to point
along the direction of the softer decay lepton in the \Znaught\ and
signal events.  Hence, we expect this opening angle to be peaked near
$\theta_{\ell\mett} = 0$ for the signal events, while for the other
backgrounds, we would not expect this.  Based on the opening angle
distributions shown in Figure~\ref{fig:lepmetangle}, we eliminate
events where $\cos\theta_{\ell\mett}<0.9$, which is particularly
effective at eliminating the \Wpart-pair background, and particularly
ineffective at eliminating signal events.

To summarize the effectiveness of our cuts, we present in
Table~\ref{table:efficiency} the fraction of each type of event which
survives each cut, along with the expected number of events
that survive; overall, roughly 40\% of the signal events will
survive all four cuts, while substantially less than 1\% of all
background events will similarly survive.

\begin{table}
\begin{center}
\newcommand{\foo}{$\sigma$ (\unit{fb})}
\newcommand{\vs}{\ensuremath{{<}0.01}}
\newlength{\ew}
\settowidth{\ew}{$100\%$}
\newcommand{\doeta}{\parbox{\ew}{\centering $\eta$}}
\begin{tabular}{|c|c|c|c|c|c|c|c|c|c|c|}
\hline
Event Type & \multicolumn{10}{c|}{Cuts}\\ \hline
& \multicolumn{2}{c|}{No Cuts} & \multicolumn{2}{c|}{\ptl} &
  \multicolumn{2}{c|}{$n_{\text{jet}}$} & \multicolumn{2}{c|}{
    $\cos\theta_{\elec\muon}$} &
  \multicolumn{2}{c|}{$\cos\theta_{\ell\mett}$}\\ \hline  
& \doeta & \foo & \doeta & \foo & \doeta & \foo &
 \doeta & \foo & \doeta & \foo \\ \hline
\multicolumn{11}{|c|}{\SUtwo\ \Zprime\ Signal, $M_\Zprime = 
\unit[450]{GeV}$}\\ \hline
$s_\phi=0.80$ & 1.00 & 35.2 & 0.80 & 28.0 & 0.67 & 23.7 & 0.58
& 20.3 & 0.50 & 17.5 \\ \hline
\multicolumn{11}{|c|}{\Uone\ \Zprime\ Signal, $M_\Zprime = \unit[350]{GeV}$}\\ 
\hline
$c_\phi=0.80$ & 1.00 & 35.7 & 0.73 & 26.0 & 0.62 & 22.3 & 0.48 &
17.2 & 0.41 & 14.7 \\ \hline
\multicolumn{11}{|c|}{Backgrounds}\\ \hline
\Znaught & 1.00 & 884.3 & \vs & 6.7 & \vs & 4.7 & \vs & 2.0 & \vs & 1.3 \\ 
Top & 1.00 & 92.7 & 0.65 & 60.0 & 0.11 & 10.6 & 0.04 & 3.9 & \vs & 0.6 \\ 
Bottom & 1.00 & 6660.0 & 0.01 & 78.0 & \vs & 2.8 & \vs & 1.7 & \vs & 0.8\\ 
\Wplus\Wminus & 1.00 & 113.4 & 0.40 & 45.6 & 0.34 & 39.1 & 0.16 & 18.4
& 0.02 & 2.7\\ 
All & 1.00 & 7750.4 & 0.02 & 190.3 & \vs & 57.2 & \vs & 26.0 & \vs & 5.4 \\ 
\hline
\end{tabular}
\end{center}
\caption{Efficiency of the four event selection cuts.  Displayed here 
  are the number of events per inverse femtobarn of luminosity of each
  type that pass each cut, along with    the cumulative efficiency,
  $\eta$, with which the given cuts retain  events.  The choice of
  where to place each of the three cuts is made to maximize the
  efficiency with which the signal is retained, while simultaneously
  maximizing the efficiency with which the backgrounds are eliminated
  and maximizing the overall number of signal events that survive.
  Overall, roughly 40-50\% of the signal events will survive, while
  less than 1\% of the total background will survive for the \ptl,
  jet, and opening angle cuts discussed in the text.} 
\label{table:efficiency}
\end{table}

After performing these cuts on our data, we determined normalized
signal and background cross sections from our Monte Carlo data, from
which we can obtain luminosity bounds for 90\% and 95\% exclusion, as
well as three and five standard deviation discovery bounds.  We will
explore first the exclusion reach of the Tevatron, followed by the
discovery reach.

Exclusion bounds are obtained by calculating the following Poisson
test statistic \cite{confidence},
\begin{equation}
r(\sigma_S, \sigma_B, \mathcal{L}) = 
1 - \frac{\sum_{i=0}^{N} 
         \left(S + B\right)^i e^{-(S+B)}/i! }{ \sum_{i=0}^{N} B^i e^{-B}/i!}
\end{equation}
where $\sigma_B$ is the calculated background cross section,
$\sigma_S$ is the calculated signal cross section, $\mathcal{L}$ is
the integrated luminosity, $B = \sigma_B \mathcal{L}$ is the expected
number of background events, $S = \sigma_S \mathcal{L}$ the expected
number of signal events, and $N$ is the largest integer smaller than
the upper limit on the expected number of background events, that is
$N = \lfloor \sigma_B \mathcal{L} \rfloor = \lfloor B \rfloor$.  For
each value of $M_{\Zprime}$ and $s_\phi$, taking the calculated signal
and background cross sections, we varied the integrated luminosity and
determined the statistic $r$.  The minimum integrated luminosity
required to exclude the model at a given confidence level,
$\confidence$, is the luminosity where $r = \confidence$.  The algorithm
determines the ratio of the total probability of $N$ or fewer events
occurring in an experiment, for the model of new physics, compared to
the probability for standard model physics.  At a given confidence
level, $\confidence$, the area of overlap between the two probability
distributions will be given by $1-\confidence$.

\begin{figure}[tb]
\begin{center}
\includegraphics[width=\textwidth-2in]{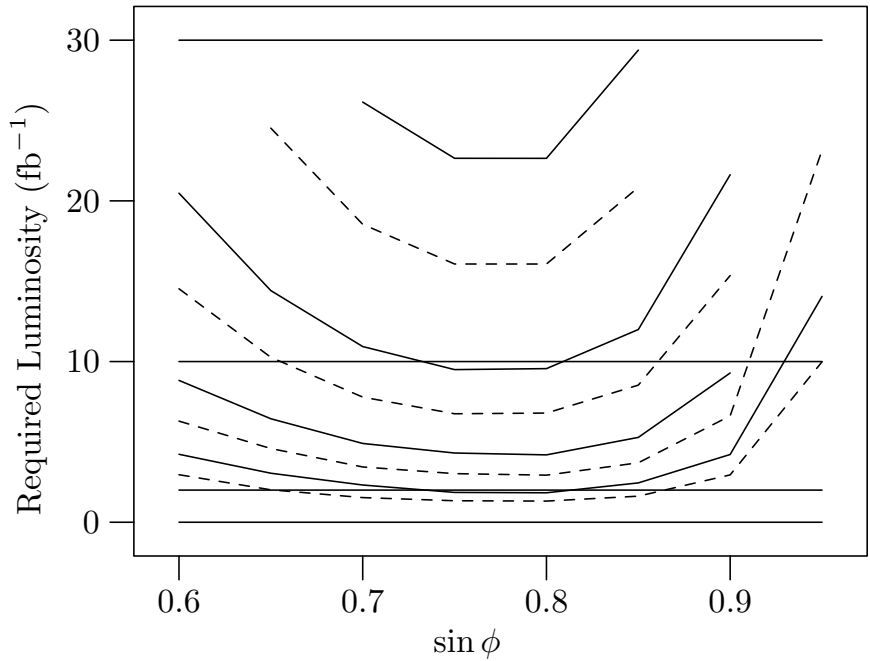}
\end{center}
\caption{Luminosity required to exclude \SUtwo{} \Zprime{} bosons
  of various mass and mixing angles in the extended electroweak
  scenario of Section~\ref{sec:ewsb}.  We display four pairs of
  curves, each with a lower dashed curve, the 90\% exclusion bound,
  and an upper solid curve, the 95\% exclusion bound.  From bottom to
  top, the curves correspond to \Zprime{} masses of $\unit[600]{GeV}$,
  $\unit[650]{GeV}$, $\unit[700]{GeV}$, and $\unit[750]{GeV}$.  The
  horizontal lines indicate luminosity targets for Run II, for
  reference: $\unit[2]{fb^{-1}}$, $\unit[10]{fb^{-1}}$, and
  $\unit[30]{fb^{-1}}$.}
\label{fig:limits-su2}
\end{figure}

We plot the exclusion limits in Figure~\ref{fig:limits-su2} for a
number of \Zprime{} masses and a range of mixing parameters, $s_\phi$.
For a \Zprime{} boson of given mass, the luminosity required for
exclusion is lowest when the mixing angle is near $s_\phi = 0.80$,
with an approximately quadratic increase on either side of the
minimum.  The shape of the exclusion curve reflects the dependence of
the \Zprime{} width on $s_\phi$ (cf. Figure~\ref{fig:widths}): narrow
\Zprime{} bosons are easier to detect. With a few inverse femtobarns
of integrated luminosity, the $\Zprime \rightarrow \tauon\tauon \to
\elec \muon$ channel will begin to explore portions of the model
parameter space that are not excluded by the precision electroweak
data.

\begin{figure}[tb]
\begin{center}
\includegraphics[width=\textwidth-2in]{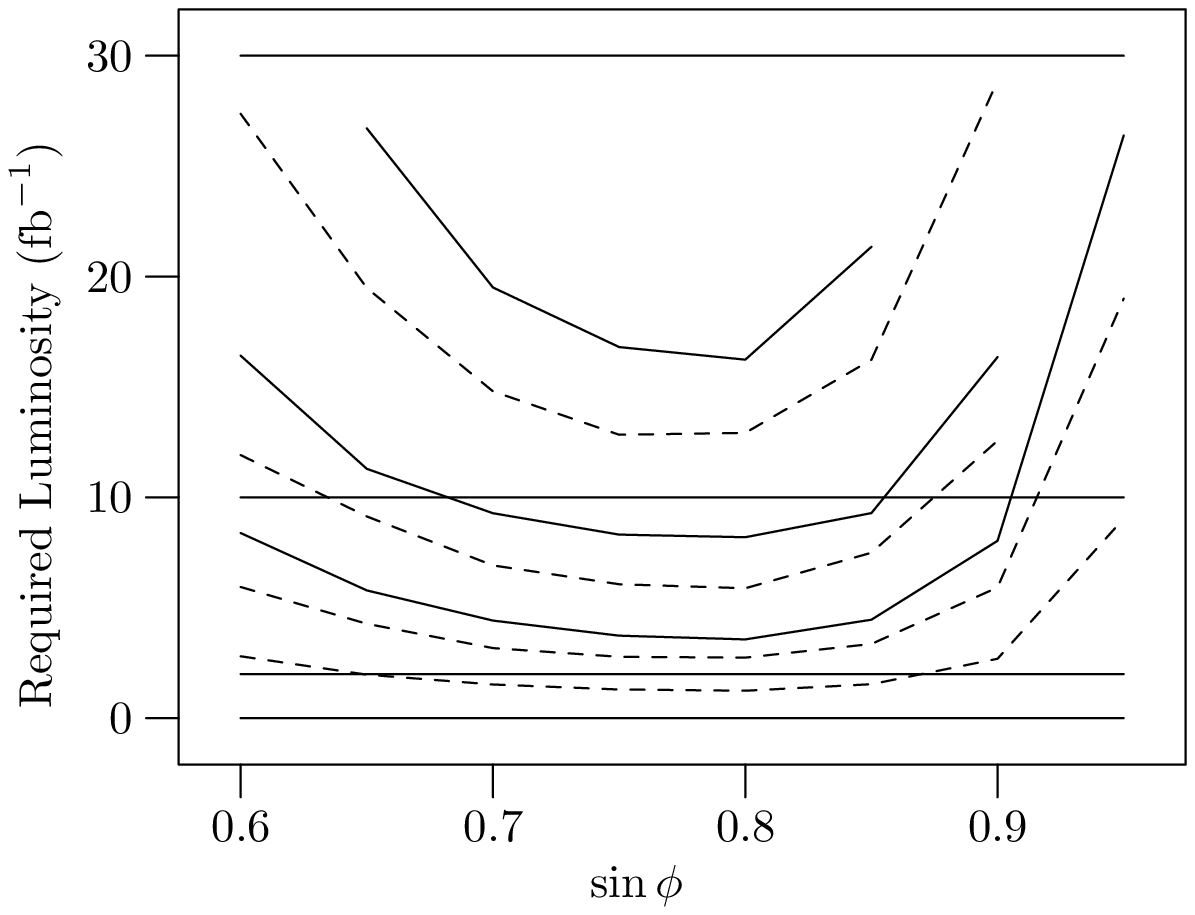}
\end{center}
\caption{Luminosity required to discover \SUtwo\ \Zprime\ bosons of
  various masses and mixing angles in the extended electroweak
  scenario of Section~\ref{sec:ewsb}.  We display two types of curves.
  Dashed curves are \threesigma\ discovery curves for a fixed mass,
  while solid curves are \fivesigma\ discovery curves.  From bottom to
  top, \threesigma\ curves are displayed for \Zprime\ masses of
  \unit[550]{GeV}, \unit[600]{GeV}, \unit[650]{GeV}, and
  \unit[700]{GeV}.  From bottom to top, \fivesigma\ curves are
  displayed for \Zprime\ masses of \unit[550]{GeV}, \unit[600]{GeV},
  and \unit[650]{GeV}.  The horizontal lines indicate luminosity
  targets for Run II, for reference: $\unit[2]{fb^{-1}}$,
  $\unit[10]{fb^{-1}}$, and $\unit[30]{fb^{-1}}$.}
\label{fig:discovery-su2}
\end{figure}

Discovery limits are obtained by applying the following algorithm.
For a given integrated luminosity, $\mathcal{L}$, we expect to observe
$B = \sigma_B \mathcal{L}$ background events, and $S = \sigma_S
\mathcal{L}$ signal events, for a total of $S+B$ expected events.  We
calculate the Poisson probability, $P(\mu, x)$, that an expected
background of $\mu = B$ events could fluctuate to give us a false
signal of $x = S+B$ events total, that is $P(B,S+B)$.  If the
probability of such a fluctuation is smaller than a given confidence
level, we are justified in declaring discovery of a new phenomenon at
that level.  We choose to determine the luminosity, $\mathcal{L}$, for
three gaussian standard deviation discovery (which we denote as
\threesigma), where $P(B, S+B)_{3\sigma} \le 1.35 \times 10^{-3}$, and
for \fivesigma, where $P(B, S+B) \le 2.7\times 10^{-7}$.  We plot
discovery bounds for a number of \Zprime\ masses and a range of mixing
parameter, $s_\phi$, in Figure~\ref{fig:discovery-su2}
As expected from the previously determined exclusion bounds, for a
\Zprime\ boson of a given mass, the luminosity for discovery is lowest
when the mixing angle is near $s_\phi = 0.80$.  
As with exclusion, only a few inverse femtobarns of integrated
luminosity will be required to discover a \Zprime\ boson with mass just
above the current exclusion bounds.

\begin{figure}[tb]
\begin{center}
  \begin{minipage}{(\textwidth-2in)/2}
    \begin{center}
      \includegraphics[width=\textwidth]{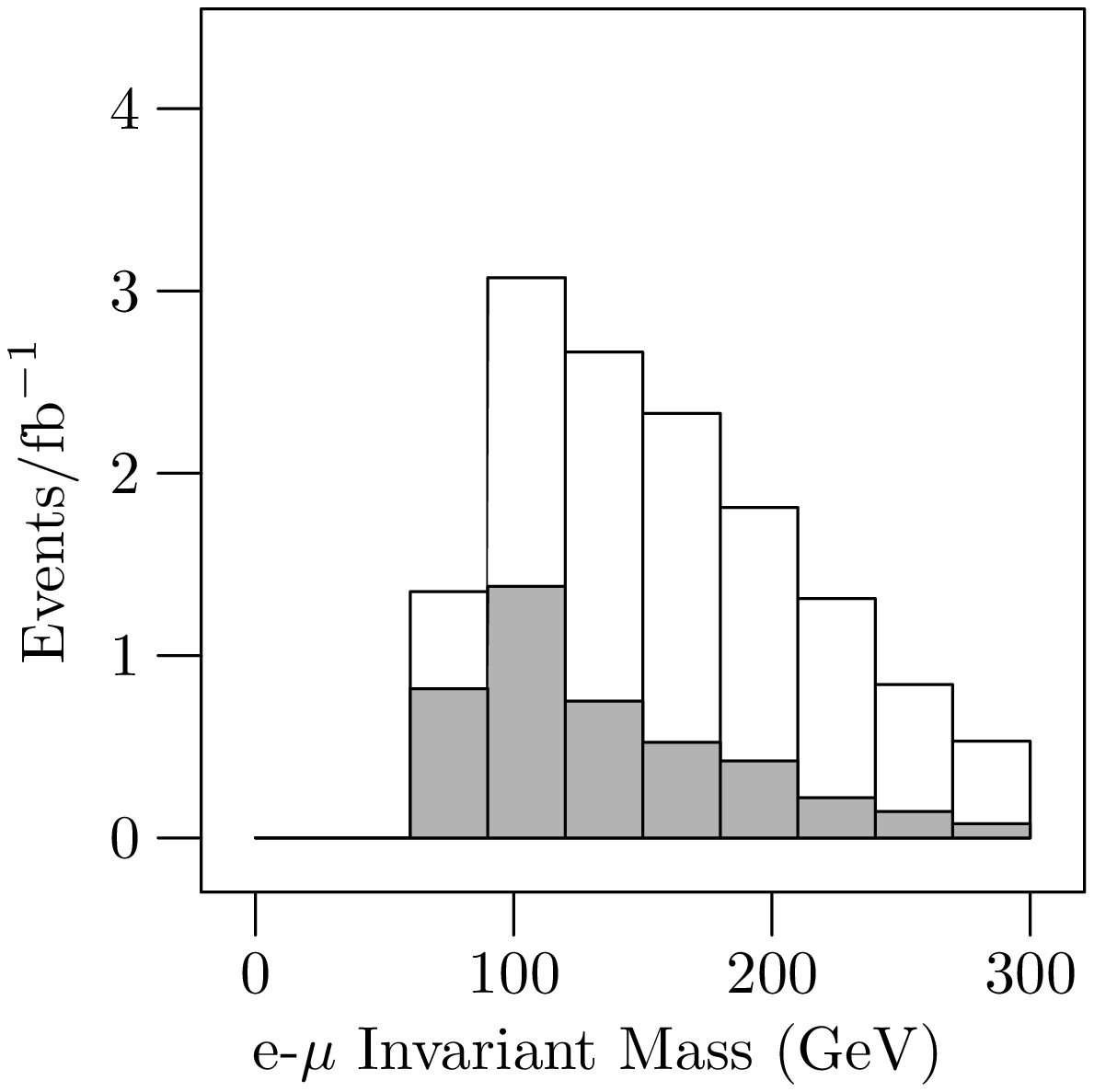}
    \end{center}
  \end{minipage}\qquad
  \begin{minipage}{(\textwidth-2in)/2}
    \begin{center}
      \includegraphics[width=\textwidth]{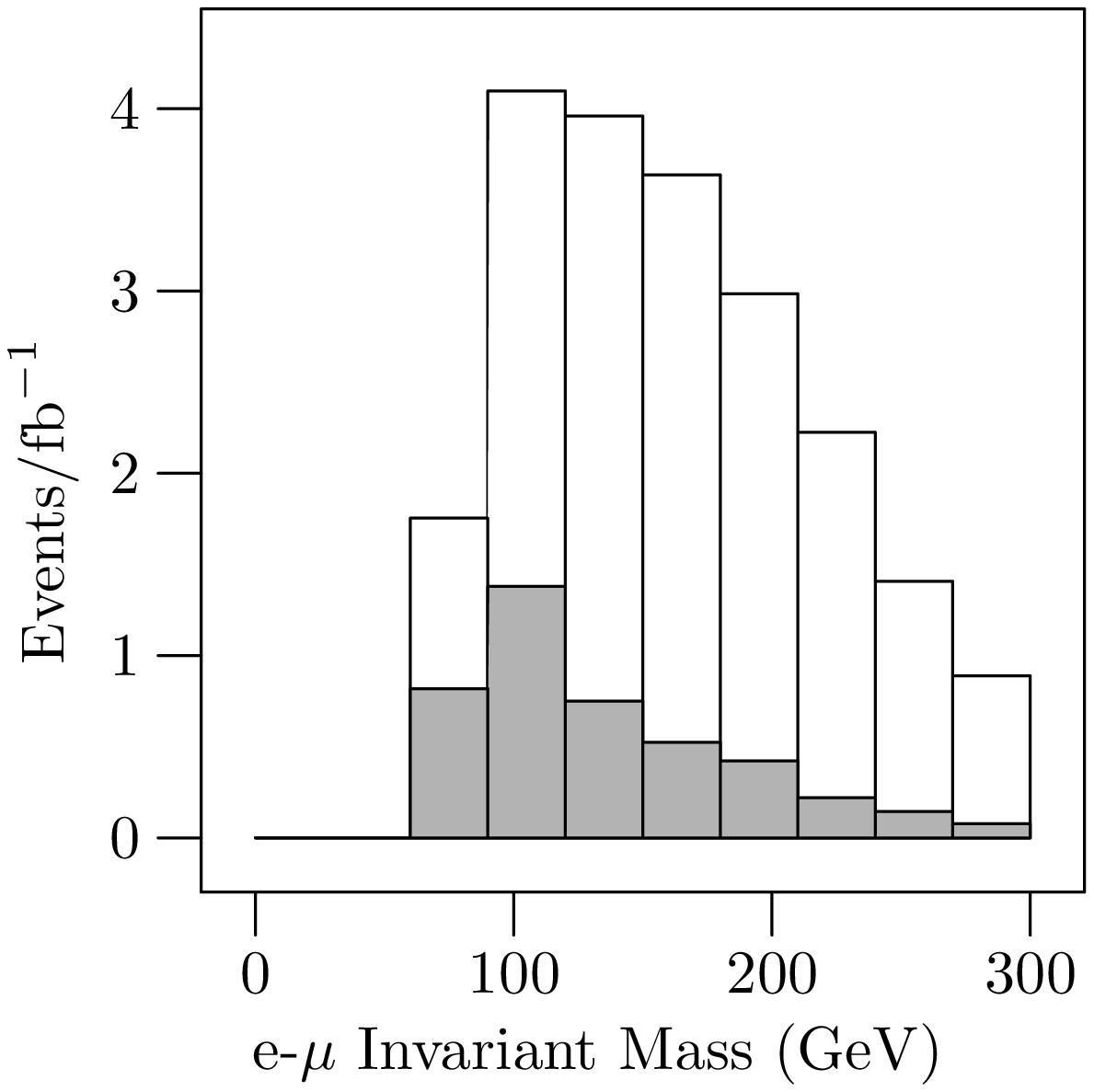}
    \end{center}
  \end{minipage}
\end{center}
\caption{Electron-muon invariant mass ($M_{\elec\muon} = |p_\elec
  + p_\muon|$) distribution for signal plus background events
  surviving all cuts.  For comparison, the distribution for all
  background events is shaded.  The presence of a heavy \Zprime{}
  boson clearly alters the distribution of events, toward higher
  invariant mass.  Variation of the mixing angle for a fixed mass
  \Zprime{} impacts the event rate of the distributions, but roughly
  speaking, not the shape or location of the peak.  The distributions
  displayed here are for $M_{\Zprime} = \unit[450]{GeV}$ and $s_\phi =
  0.60$, and for $M_{\Zprime} = \unit[450]{GeV}$ and $s_\phi = 0.80$.}
\label{fig:inv-mass-1}
\end{figure}

\begin{figure}[tb]
\begin{center}
  \begin{minipage}{(\textwidth-2in)/2}
    \begin{center}
      \includegraphics[width=\textwidth]{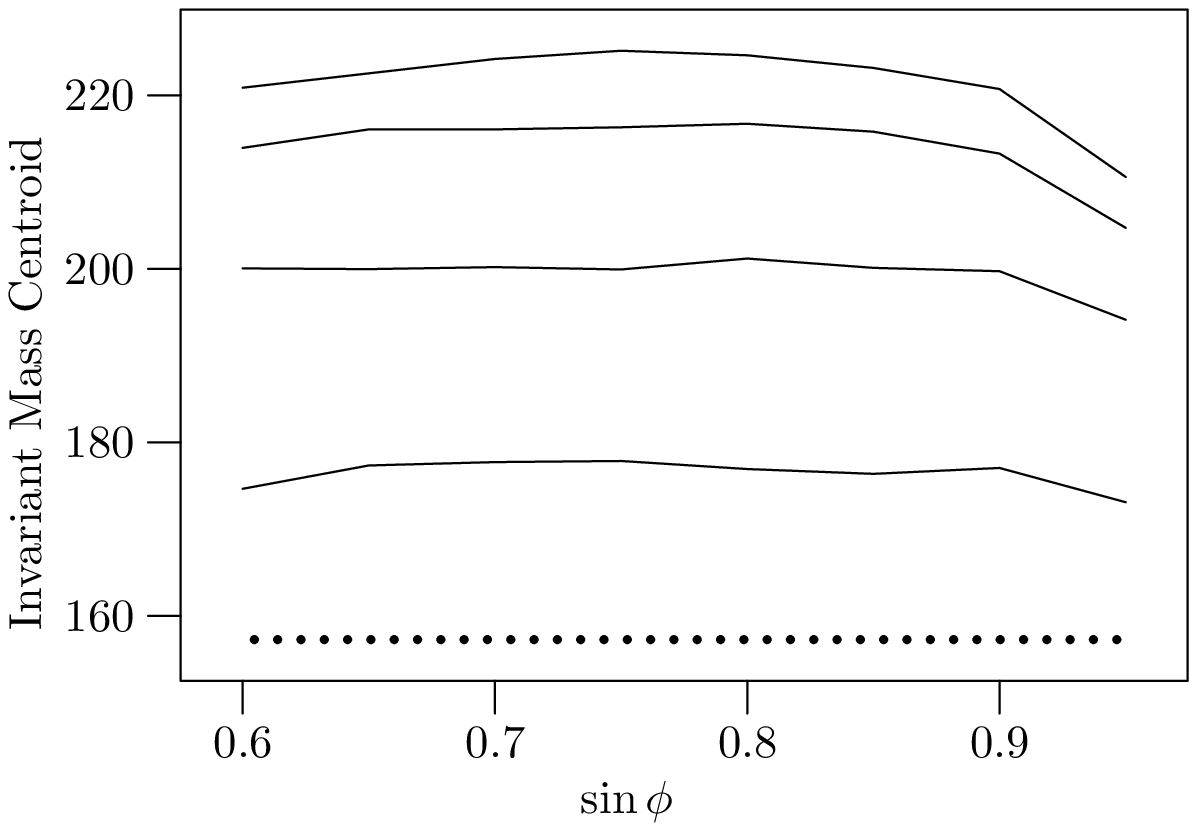}
    \end{center}
  \end{minipage}\qquad
  \begin{minipage}{(\textwidth-2in)/2}
    \begin{center}
      \includegraphics[width=\textwidth]{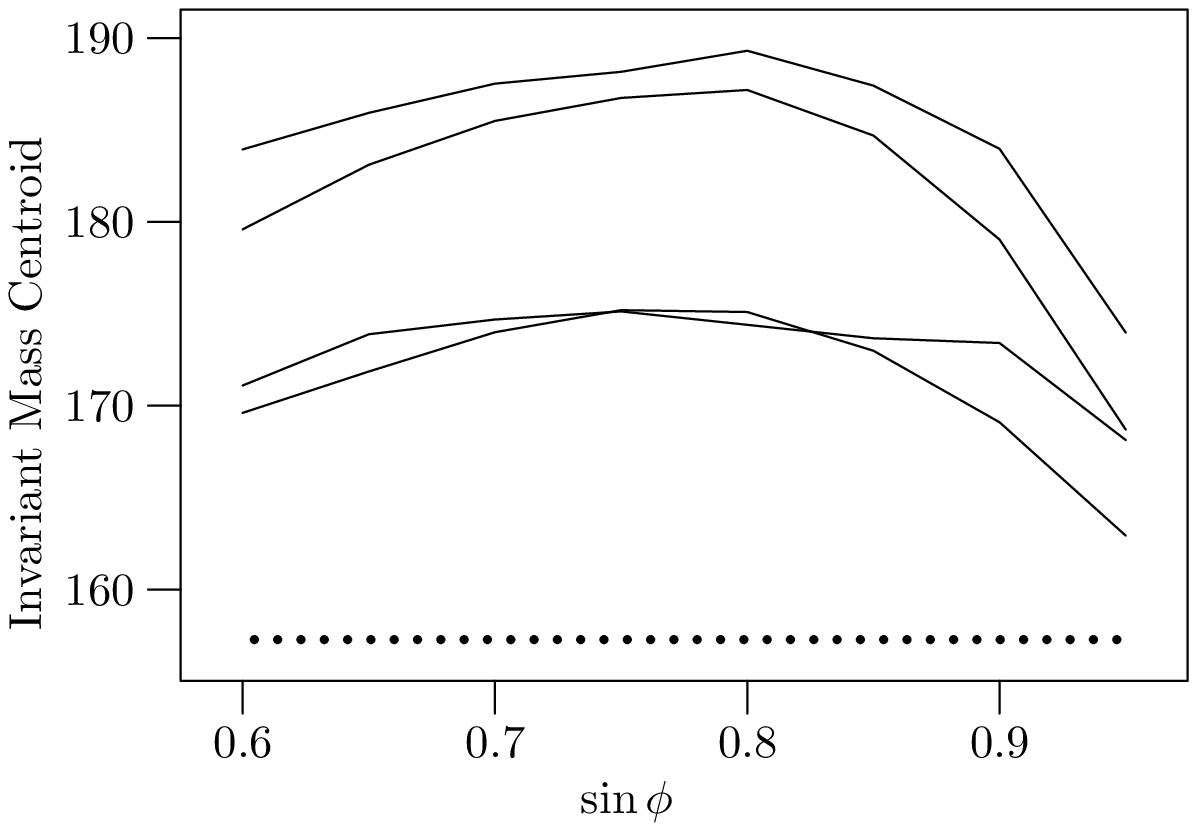}
    \end{center}
  \end{minipage}
\end{center}
\caption{We plot the centroids of the \elec\muon\ invariant mass
  distributions, $M_{\elec\muon}$.  The plot on the left displays the
  centroids of the signal distributions as a function of mass and
  mixing angle.  Note that the centroids are nearly independent of the
  mixing angle.  We do not, however, measure the signal in isolation,
  but in the presence of backgrounds.  The plot on the right displays
  the centroids of the combined signal and background distributions.
  Here, the separation of the model parameters based on the data is
  less straightforward, although clearly differentiated from the
  background distribution alone.  The solid curves, from bottom to top
  in the left plot, correspond to \Zprime{} masses of \unit[450]{GeV},
  \unit[550]{GeV}, \unit[650]{GeV}, and \unit[750]{GeV}; the dotted
  line at the bottom is the centroid of the background by itself.  The
  signal plus background distributions in the right plot all
  essentially overlap, but are separated from the background
  distribution.  The dotted line at the bottom is the centroid of the
  background by itself.  The solid curves, from bottom to top,
  correspond to \Zprime\ masses of \unit[750]{GeV}, \unit[450]{GeV},
  \unit[650]{GeV}, and \unit[550]{GeV}.}
\label{fig:centroid}
\end{figure}

By studying the \elec\muon\ invariant mass distribution
($M_{\elec\muon} = | p_\muon + p_\elec |$) it may be
possible to detect the presence of a \Zprime\ boson and to determine
its mass.  As shown in Figure~\ref{fig:inv-mass-1}, 
the invariant mass distribution of the background events which pass
all of our cuts peak at around $\unit[100]{GeV}$.  The centroid of the
distribution of signal events is shifted toward higher invariant mass,
with the amount of the shift depending on the value of $M_\Zprime$,
but almost independent of $s_\phi$, Figure~\ref{fig:centroid}.  
However, as the background rate is independent of the mixing angle
while the signal rate is not, the centroid of the background plus
signal distribution is not sufficient on its own to determine the
mass.  With a sufficiently high data rate, available at the \LHC{} for
example, it may be possible to choose more aggressive cuts that would
lessen the dependence of the centroid on the background; it may
alternatively be possible to perform a background subtraction on the
invariant mass distribution, although this approach is more difficult
to perform with confidence.  With enough data, then, it should be
possible to determine $M_\Zprime$ from the measured invariant mass
distribution $M_{\elec\muon}$.

Determining the value of $s_\phi$ will be more challenging.  As
indicated by the shape of the exclusion curves in
Figure~\ref{fig:limits-su2}, the relationship between event-rate and
$s_\phi$ is double-valued: a given event rate corresponds to two
values of $s_\phi$, one above and one below $s_\phi = 0.80$ (approx.).
Finding evidence of $\Zprime \rightarrow \eplus \eminus,\ 
\muplus\muminus$ may allow one to differentiate between the two
possible values of $s_\phi$ due to the different forms of the
couplings:  the first or second generation lepton couplings vary as
$c_\phi / s_\phi$, while to the third generation couplings vary as
$s_\phi / c_\phi$.

\section{\Zprime{} Bosons from Extended Hypercharge Interactions}
\label{sec:exthyp}

Models with an extended hypercharge gauge group can also produce heavy
\Zprime{} bosons that couple more strongly to the third generation
than to the lighter generations \cite{topassist, topastmod, bbhk,
zpcol}.  In these theories, the electroweak gauge group is
\begin{equation}
\SUtwo_W \times \Uone_h \times \Uone_\ell
\end{equation}
where third-generation fermions couple to $\Uone_h$ with standard
hypercharge values and the other fermions are carry standard
hypercharges under $\Uone_\ell$.  At a scale above the weak scale, the
two hypercharge groups break to their diagonal subgroup, identified as 
$\Uone_Y$.
As a result, a \Zprime{} boson that is a linear combination of the
original two hypercharge bosons becomes massive.  This heavy \Zprime{}
boson couples to fermions as
\begin{equation} 
-i\frac{e}{\cos\theta} \left( \frac{s_\chi}{c_\chi} Y_\ell 
-\frac{c_\chi}{s_\chi} Y_h \right)
\label{eq:zpru1}
\end{equation}
where $\chi$ is the mixing angle between the two original hypercharge
sectors
\begin{equation}
\cot{\chi} = \left(\frac{g_h}{g_\ell}\right)^2\ .
\end{equation}
Comparing Equation~\ref{eq:zpru1} with the covariant derivative for
the \Zprime{} boson from an extended weak group,
Equation~\ref{eqn:covariant_b}, we find three key differences.  Two
are physically relevant: the overall coupling is of hypercharge rather
than weak strength, and the \Zprime{} couples to both left-handed and
right-handed fermions at leading order. One is a matter of convention:
mixing angle $\chi$ is equivalent to $\pi/2 - \phi$.

At energies well below the mass of the \Zprime{} boson, its exchange
in the process $\eplus \eminus \rightarrow f \Dbar{f}$ where $f$ is a
\tauon{} lepton or \qbottom{} quark may be approximated by the contact
interaction
\begin{equation} 
\LNC \supset \frac{e^2}{\cos^2\theta M^2_{\Zprime}}
\left(\frac{s_\chi}{c_\chi} \big[\Dbar{\elec} \gamma_\mu Y_\ell
  \elec\big]\right) \left(\frac{c_\chi}{s_\chi} \big[\Dbar{f}
  \gamma^\mu Y_h f\big]\right)  
\end{equation}

Comparing this with the contact interactions studied by \LEP{} (see
Section~\ref{sec:lepdata}), we find that the \LEP{} data sets its
strongest limit through the process $\eplus_R \eminus_R \rightarrow \tauplus_R
\tauminus_R$ \cite{aleph-183, opal-183},
\begin{equation}
\Lambda(f = \tauon,\ \eta_{RR} = +1) >\begin{cases} \unit[3.7]{TeV} &
  \text{\ALEPH}\\ \unit[3.7]{TeV} & \text{\OPAL}\ ,\end{cases}
\end{equation}
which gives a limit on the \Zprime\ mass of
\begin{equation} 
M_{\Zprime} = \Lambda \sqrt{\frac{\alphaem}{\cos^2\theta}} 
          > \unit[370]{GeV}\ .
\end{equation}
This is stronger than the previous limits from precision electroweak
data \cite{rscjt}.

\begin{figure}[tb]
\begin{center}
\includegraphics[width=\textwidth-2in]{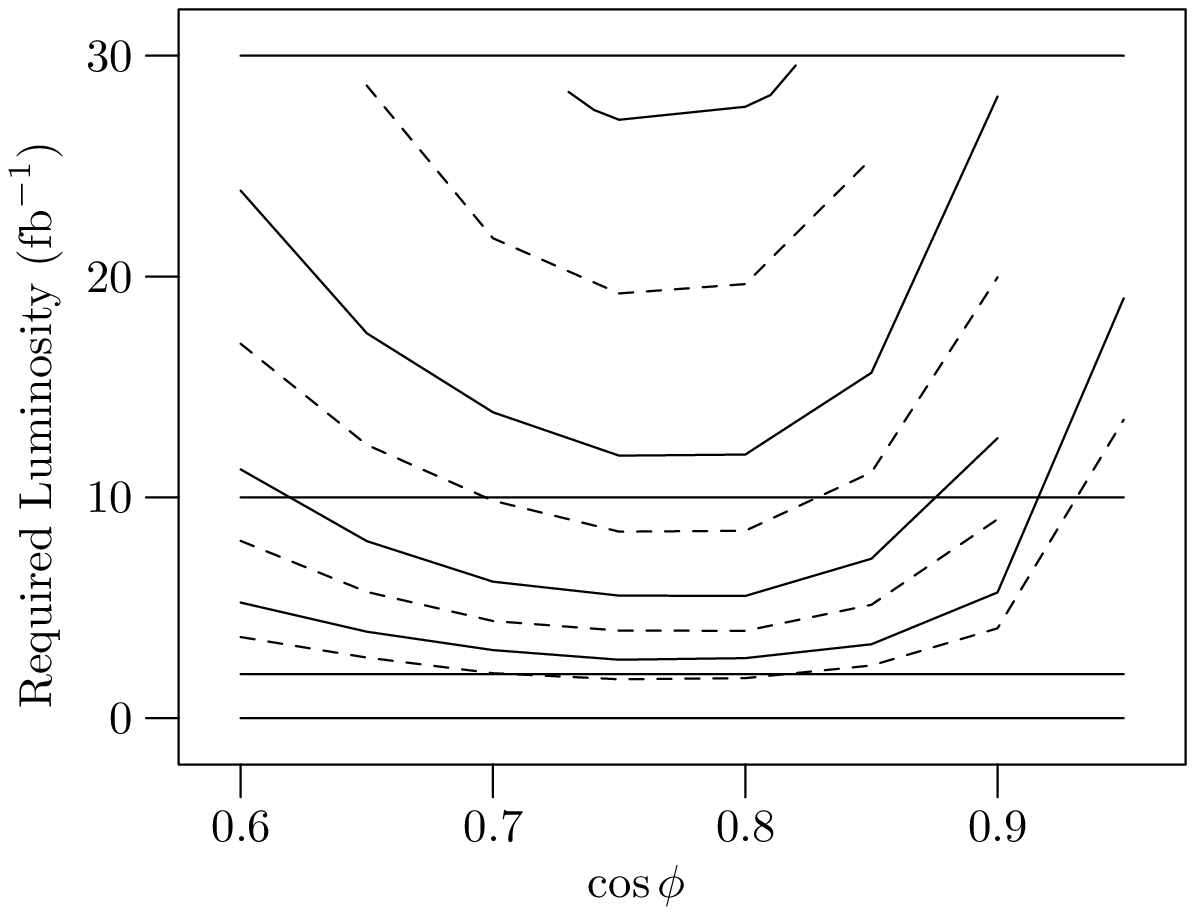}
\end{center}
\caption{Luminosity required to exclude \Uone\ \Zprime\ bosons
  of various mass and mixing angles in the extended hypercharge
  scenario of Section~\ref{sec:exthyp}.  We display four pairs of
  curves, each with a lower dashed curve, the 90\% exclusion bound,
  and an upper solid curve, the 95\% exclusion bound.  From bottom to
  top, the curves correspond to \Zprime{} masses of
  $\unit[500]{GeV}$, $\unit[550]{GeV}$, $\unit[600]{GeV}$, and
  $\unit[650]{GeV}$.  The horizontal lines correspond to the
  luminosity targets for Run II, displayed for ease of reference:
  $\unit[2]{fb^{-1}}$, $\unit[10]{fb^{-1}}$, and
  $\unit[30]{fb^{-1}}$.}
\label{fig:limits-u1}
\end{figure}

\begin{figure}[tb]
\begin{center}
\includegraphics[width=\textwidth-2in]{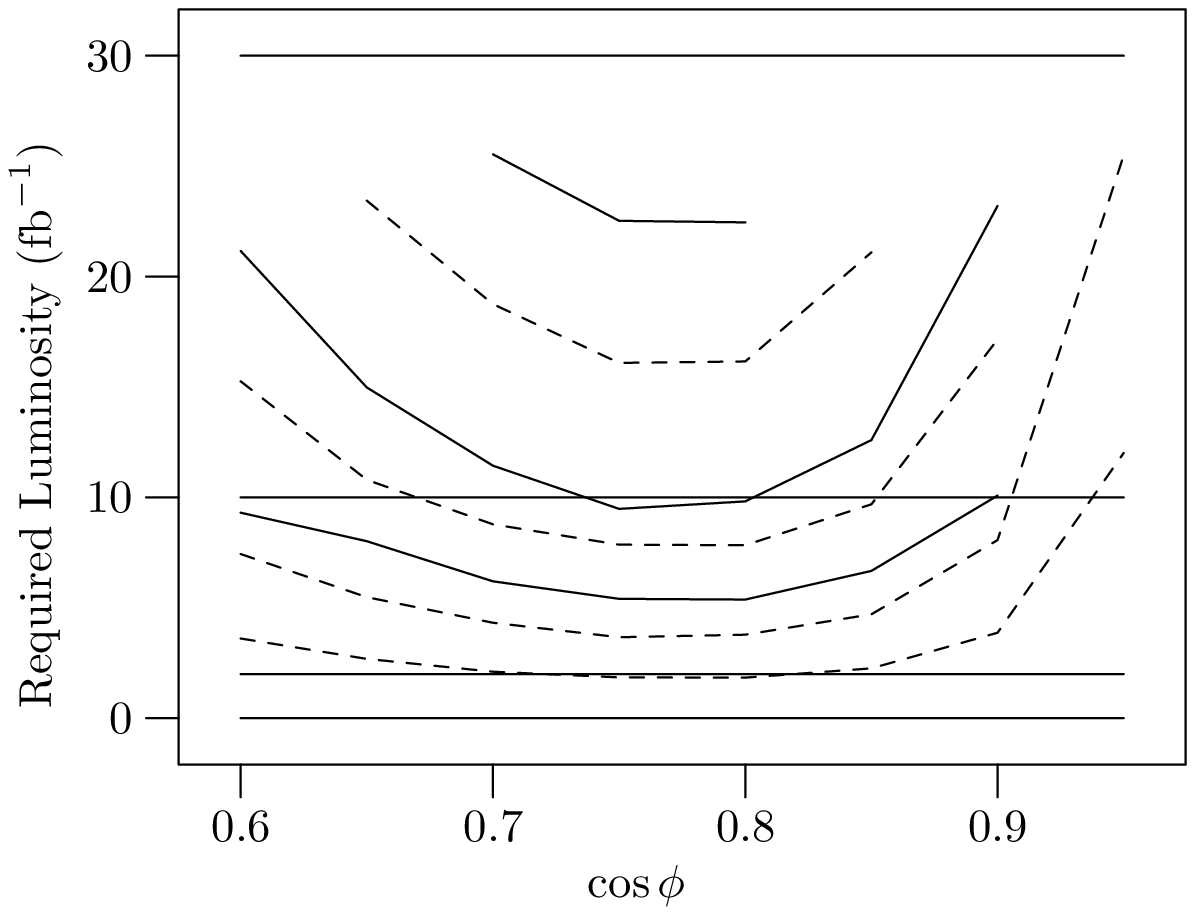}
\end{center}
\caption{Luminosity required to discover \Uone\ \Zprime\ bosons of
  various masses and mixing angles in the extended hypercharge
  scenario of Section~\ref{sec:exthyp}.  We display two types of curves.
  Dashed curves are \threesigma\ discovery curves for a fixed mass,
  while solid curves are \fivesigma\ discovery curves.  From bottom to
  top, \threesigma\ curves are displayed for \Zprime\ masses of
  \unit[450]{GeV}, \unit[500]{GeV}, \unit[550]{GeV}, and
  \unit[600]{GeV}.  From bottom to top, \fivesigma\ curves are
  displayed for \Zprime\ masses of \unit[450]{GeV}, \unit[500]{GeV},
  and \unit[550]{GeV}.  The horizontal lines indicate luminosity targets
  for Run II, for reference: $\unit[2]{fb^{-1}}$,
  $\unit[10]{fb^{-1}}$, and $\unit[30]{fb^{-1}}$.}
\label{fig:discovery-u1}
\end{figure}

We have also used the techniques described in
Section~\ref{sec:channels} to analyze the process
$\proton\Dbar{\proton} \rightarrow \Zprime \rightarrow \tauon\tauon
\rightarrow \elec \muon$ for a \Uone{} \Zprime{} boson.  Due to the
similar form of couplings of the \Zprime{} bosons to fermions, we
obtain results for exclusion and discovery bounds that can be expected
from the Tevatron that depend on mixing angle in a similar fashion.
We display exclusion bounds in Figure~\ref{fig:limits-u1}
and discovery bounds in Figure~\ref{fig:discovery-u1}.  
The luminosity required to exclude or discover a \Uone{} \Zprime{}
boson is a bit greater than for an \SUtwo{} \Zprime{} boson of the
same mass.  This difference reflects the fact that the \Uone{} boson's
coupling to fermions is of hypercharge rather than weak strength.


\section{Future Searches}
\label{sec:disc}

Detecting even relatively light \Zprime{} bosons that couple
preferentially to third-generation fermions is clearly a challenge for
Tevatron and LEP experiments.  Even in the $p\bar{p} \rightarrow
\Zprime \rightarrow \tauon\tauon \rightarrow \elec\muon$ process where
the signal-to-background ratio can be made quite large, the absolute
number of signal events is kept low by the size of the \Zprime{}
boson's coupling to the light fermions from which it is produced.  In
the long term, the \LHC 's higher center-of-mass energy will allow its
experiments to search for these \Zprime{} bosons without being
hampered by low signal event rates.  In the meantime, we suggest that
a few additional search channels may prove useful.

Obviously, the reach in \tautau\ final states will be extended beyond
that shown in this analysis if use can be made of one or more hadronic
tau decays \footnote{A parton-level study in ref. \cite{dproy}
estimates that a 500 GeV X boson coupling to $B - 3 L_\tau$ could be
visible via $X \rightarrow \tautau \rightarrow \text{jet} + \ell$ in 2
$fb^{-1}$ of data at Run II.}.  The single-prong decays of the tau,
which constitutes about 85\% of all decays, may have sufficiently
small background since QCD should rarely produce isolated, high-\pt\
tracks.  It will be difficult to use these final states to search for
\Zprime\ bosons: it is extremely hard, in the hadronic environment, to
trigger on jets, and flavor tagging with high precision is an
unresolved problem. Nonetheless, since the branching ratio of \tauon{}
to hadrons ($\mathrm{BR}(\tauon \rightarrow \text{hadrons}) \approx
65\%$) is higher than to leptons ($\mathrm{BR}(\tauon \rightarrow
\text{leptons}) \approx 35\%$) \cite{pdg}, even modest jet trigger and
flavor tagging efficiencies could prove extremely valuable in searches
or measurement of parameters.  The ability to use these additional
channels with their higher event rates should yield significantly
improved mass limits for a given integrated luminosity.

In the semi-leptonic decay scenario, where we have $\tautau
\rightarrow \text{jet} + \ell$, the event trigger could be a high-\pt\
electron or muon with, for example, $\pt > \unit[15]{GeV}$.  In
offline processing, one would then reconstruct the jets, and attempt
to perform flavor tagging.  If the corrected \tauon\ tagging
efficiency\footnote{By which we mean the tagging efficiency, after
corrections for other objects faking \tauon\ jets.}  can be raised to
approximately 15\%, then the semi-leptonic events will provide the
same event rate for analysis as the fully leptonic events previously
considered.   Further study of these channels are clearly
warranted.

\Zprime{} bosons arising from extended weak interactions will also be
accompanied by \Wprime{} bosons of very similar mass (to leading
order, $M_\Zprime = M_\Wprime$).  These bosons could be searched for in
the process $\proton\Dbar{\proton} \rightarrow \Wprime \rightarrow
\tauon \neutrino_\tauon$.  Standard model backgrounds would include
$\proton\Dbar{\proton} \rightarrow W \rightarrow \ell \neutrino_\ell$
and $\proton\Dbar{\proton} \rightarrow \Wpart \Znaught \rightarrow
\ell \neutrino_\ell \neutrino \neutrino$, both of which should have
softer lepton spectra, as well as $\proton\Dbar{\proton} \rightarrow
\Znaught + \text{jet} \rightarrow \neutrino \neutrino + \text{fake
  lepton}$, where the jet is misidentified.

The methods of analysis pursued in Section~\ref{sec:channels} could
productively be applied to models with scalars that have large
branching ratios to tau pairs. While a heavy Standard Model Higgs
boson does not have a high enough branching ratio for these analyses
to provide useful limits, a pseudoscalar Higgs with large branching
ratio would be an interesting candidate for study.  


\section{Conclusions}
\label{sec:conclusions}

We have discussed two methods of searching for \Zprime\ bosons that
couple primarily to third generation fermions.  Bounds on the scale of
quark-lepton compositeness derived from data taken at \LEP\ and the
Tevatron now imply that \Zprime\ bosons derived from extended
$\SUtwo_h\times \SUtwo_\ell$ or $\Uone_h\times \Uone_\ell$
interactions must have a mass greater than about $\unit[375]{GeV}$.
The reach of these limits will improve as additional data is taken.
As the Tevatron Run II begins, it will become possible to search for
\Zprime\ bosons using the process $p\overline{p} \to \Zprime \to
\tauon\tauon \to \elec\muon \particle{X}$.  We have shown that a
combination of cuts based on lepton transverse momenta, jet
multiplicity, and event topology, can overcome the standard model
backgrounds.  With $\unit[30]{fb^{-1}}$ of data, the Run II
experiments will be able to exclude \Zprime\ bosons with masses up to
$\unit[750]{GeV}$.  Were a \Zprime\ boson, instead, discovered, the
shape of the \elec\muon\ invariant mass distribution and the relative
branching fractions to taus and to muons could reveal the \Zprime\ 
mass and coupling strength.

\vspace{24pt} 
\begin{center}\textbf{Acknowledgments}\end{center} 
\vspace{12pt} 

The authors thank C.~Hoelbling, F.~Paige, and M.~Popovic for useful
conversations, and K.~Lane and P.~Kalyniak for comments on the
manuscript.  E.H.S.  acknowledges the support of the NSF Faculty Early
Career Development (CAREER) program and the DOE Outstanding Junior
Investigator program.  M.N. acknowledges the support of the NSF
Professional Opportunities for Women in Research and Education (POWRE)
program.  {\em This work was supported in part by the National Science
  Foundation under grants PHY-9501249 and PHY-9870552, by the
  Department of Energy under grant DE-FG02-91ER40676, and by the Davis
  Institute for High Energy Physics.}

\newcommand{\npb}{\textit{ Nucl.\ Phys.}\ \textbf{ B}} 
\newcommand{\pr}{\textit{ Phys.\ Rev.}\ } 
\newcommand{\prd}{\textit{ Phys.\ Rev.}\ \textbf{ D}} 
\newcommand{\prp}{\textit{ Phys.\ Rep.}\ } 
\newcommand{\prl}{\textit{ Phys.\ Rev.\ Lett.}\ } 
\newcommand{\pl}{\textit{ Phys.\ Lett.}\ \textbf{ B}} 
\newcommand{\ptp}{\textit{ Prog.\ Theor.\ Phys.}\ } 
\newcommand{\ap}{\textit{ Ann.\ Phys.}\ } 
\newcommand{\intl}{\textit{ Int.\ J.\ Mod.\ Phys.}\ \textbf{ A}} 
\newcommand{\mpl}{\textit{ Mod.\ Phys.\ Lett.}\ \textbf{ A}} 
\newcommand{\cpc}{\textit{Comp. Phys. Comm.}} 
\newcommand{\epjc}{\textit{Eur.\ Phys.\ J.}\ \textbf{C}}

 
\end{document}